\documentclass[twocolumn]{aastex63}
\usepackage{amsmath,amsopn}
\usepackage{comment}
\usepackage{natbib} 
\usepackage{epstopdf}
\usepackage{tabularx}
\usepackage{verbatim}
\usepackage{booktabs}
\usepackage{nicefrac}
\usepackage{graphicx}

\newcommand{\kms}{{\rm km}~{\rm s}^{-1}}
\newcommand{\kpc}{\ensuremath{\, \mathrm{kpc}}}
\newcommand{\cm}{\ensuremath{ \, \mathrm{cm}}}
\newcommand{\hi}{H\textsc{~i}}
\newcommand{\hicm}{H\textsc{~i}~21-cm}

\newcommand{\ha}{H$\alpha$}
\newcommand{\angstrom}{\mbox{\normalfont\AA}}
\newcommand{\sightlines}{1,712 }

\submitjournal{ApJ}

\shorttitle{LMC Galactic Outflow}
\shortauthors{Ciampa et al.}

\begin{document}

\title{Mapping the Supernovae Driven Winds of the Large Magellanic Cloud in \ha\ Emission I}

\email{drew.ciampa@tcu.edu }

\author[0000-0002-1295-988X]{Drew A. Ciampa}
\affiliation{Department of Physics \& Astronomy, Texas Christian University, Fort Worth, TX 76129, USA}

\author[0000-0001-5817-0932]{Kathleen A. Barger}
\affiliation{Department of Physics \& Astronomy, Texas Christian University, Fort Worth, TX 76129, USA}

\author{Nicolas Lehner}
\affiliation{Department of Physics, University of Notre Dame, Notre Dame, IN 46556, USA}

\author{Madeline Horn}
\affiliation{Department of Physics \& Astronomy, Texas Christian University, Fort Worth, TX 76129, USA}
\affiliation{Department of Astronomy, Smith College, Northampton, MA 01063, USA}

\author{Michael Hernandez}
\affiliation{Department of Physics \& Astronomy, Texas Christian University, Fort Worth, TX 76129, USA}

\author{L. Matthew Haffner}
\affiliation{Embry-Riddle Aeronautical University, Daytona Beach, FL 32114, USA}
\affiliation{Space Science Institute, Boulder, CO 80301, USA}
\affiliation{Department of Astronomy, University of Wisconsin-Madison, Madison, WI 53706, USA}

\author{Brianna Smart}
\affiliation{Department of Physics, Astronomy, and Mathematics, University of Hertfordshire, Hatfield AL10 9AB, UK}
\affiliation{Department of Astronomy, University of Wisconsin-Madison, Madison, WI 53706, USA}

\author[0000-0002-8366-2143]{Chad Bustard}
\affiliation{Department of Physics, University of Wisconsin-Madison, Madison, WI 53706, USA}

\author{Sam Barber}
\affiliation{Department of Physics \& Astronomy, Texas Christian University, Fort Worth, TX 76129, USA}
\affiliation{Trinity Valley High School, Fort Worth, TX 76132, USA}

\author{Henry Boot}
\affiliation{Department of Physics \& Astronomy, Texas Christian University, Fort Worth, TX 76129, USA}
\affiliation{Burleson High School, Burleson, TX 76028, USA}

\begin{abstract}
We present the first spectroscopically resolved \ha\ emission map of the Large Magellanic Cloud's (LMC) galactic wind. By combining new Wisconsin H-alpha Mapper (WHAM) observations ($I_{\rm H\alpha}\gtrsim10~{\rm mR}$) with existing \hicm\ emission observations, we have (1) mapped the LMC's near-side galactic wind over a local standard of rest (LSR) velocity range of $+50\le\rm v_{LSR}\le+250~\kms$, (2) determined its morphology and extent, and (3) estimated its mass, outflow rate, and mass-loading factor. We observe \ha\ emission from this wind to typically 1-degree off the LMC's \hi\ disk. Kinematically, we find that the diffuse gas in the warm-ionized phase of this wind persists at both low ($\lesssim100~\kms$) and high ($\gtrsim100~\kms$) velocities, relative to the LMC's \hi\ disk. Furthermore, we find that the high-velocity component spatially aligns with the most intense star-forming region, 30~Doradus. We, therefore, conclude that this high-velocity material traces an active outflow. We estimate the mass of the warm ($T_e\approx10^4~\rm K$) ionized phase of the near-side LMC outflow to be $\log{\left(M_{\rm ionized}/M_\odot\right)=7.51\pm0.15}$ for the combined low and high velocity components. Assuming an ionization fraction of 75\% and that the wind is symmetrical about the LMC disk, we estimate that its total (neutral and ionized) mass is $\log{\left(M_{\rm total}/M_\odot\right)=7.93}$, its mass-flow rate is $\dot{M}_{\rm outflow}\approx1.43~M_\odot~\rm yr^{-1}$, and its mass-loading factor is $\eta\approx4.54$. Our average mass-loading factor results are roughly a factor of 2.5 larger than previous \ha\ imaging and UV~absorption line studies, suggesting that those studies are missing nearly half the gas in the outflows.
\end{abstract}


\section{Introduction} \label{sec:intro}

Galactic feedback, such as stellar winds, supernovae, and active galactic nuclei, expel both energy and momentum into the interstellar medium (ISM) of the host galaxy. These processes can further drive gas out of the galaxies in galactic winds and fountains. As these processes cycle gaseous material through the galaxy and into its surroundings, they transport enriched gas to the outskirts of the galaxy and into the circumgalactic medium (CGM; \citealt{heckman2003}, \citealt{veilleux2005}, \citealt{Tumlinson2011}). Furthermore, if the gas that is ejected into the CGM is lost from the galaxy or if it stagnates into the galaxy's halo (e.g., \citealt{ford2014, peeples2014}), the star-formation rate of the galaxy will likely decline unless it is able to procure additional gaseous material from different sources. In most cases, this baryon cycle is difficult to resolve because the gaseous material in the ISM and CGM is faint. However, the nearby Large Magellanic Cloud (LMC) provides an unparalleled view of gaseous material both within and surrounding its disk. By investigating the gaseous material in the CGM of the LMC, we can better understand its behavior, enabling us to decipher how the baryon cycle is connected to galaxy evolution.

At a distance of about $d_\odot\approx50~\kpc$ \citep{Pietrzy2013,deGrijs2014,Walker1999}, the LMC is close enough to resolve spatial features within its ISM. The stellar and total mass, $M_\star=3\times10^9~M_\odot$ \citep{vandermarel2009} and $M_{\rm total}=1.7\times10^{10}~M_\odot$ ($\rm r_{\rm enclosed}=8.7~\kpc$; \citealt{vandermarel2014}), make the LMC a low-mass galaxy allowing gas to interact more freely with its environment and distribute material with the CGM efficiently \citep{Heckman2000}. The gaseous disk of this galaxy is projected nearly face-on with an inclination angle of $22\arcdeg\lesssim i\lesssim26\arcdeg$ \citep{Kim1998,Staveley2003,Choi2018}, providing an unobstructed view of its interstellar medium and activity within the disk. Observations have shown numerous neutral hydrogen super shells and holes that exist throughout the LMC's \citep{Meaburn1980,Kim1999}, which could be a result of interactions with the Small Magellanic Cloud (SMC) and possibly the Milky Way (MW; e.g., \citealt{Besla2010, Besla2012}), as well as recent periods of intense star formation due to its interaction with the SMC \citep{Harris2009}. Each of these studies suggests an active history and stellar lifecycle that could lead to outflows with large amounts of energy and material blown into their surroundings during times of increased stellar activity \citep{Erb2015}. The addition of energy and momentum from numerous supernovae throughout the galaxy could be a way for a large-scale outflow to originate in a galaxy like the LMC. Observations capturing a complete picture of any galactic-wide outflowing material proved difficult due to the LMC's size on the sky. Observations of the \ha\ emission from the ISM of the LMC in prior studies focused primarily inside the LMC's disk (\citealt{Rosado1990}, \citealt{Laval1992}, and \citealt{Reid2012}). While these studies revealed activity within the ISM of the LMC (\citealt{pellegrini2012} and \citealt{wink2015}), they only observed the brighter inner region of the LMC rather than the faint emission from its extended diffuse disk and the galaxy's circumgalactic medium.

Although the LMC is our nearest gas-rich neighboring galaxy, it was not until recently that many studies began to directly detect signatures of a large-scale galactic outflow. Recent ultraviolet (UV) absorption-line spectroscopy studies, using the \textit{Hubble Space Telescope} (\textit{HST}) and \textit{Far Ultraviolet Spectroscopic Explorer} (\textit{FUSE}), investigated gas flows of the LMC \citep{Howk2002,Lehner2007,Lehner2009, Pathak2011}. \citet{Howk2002} used 12~stars embedded within the LMC as background targets to explore the gas on the near-side of the LMC; they found that the absorption along every sightline had kinematic signatures consistent with gaseous material flowing out of the LMC. A study performed by \citet{Staveley2003} using the \hicm\ emission-line surveyed found high-velocity clouds in the direction of the LMC and kinematic and morphological evidence that these clouds could be associated with an LMC galactic outflow. A subsequent absorption-line investigation conducted by \citet{Lehner2007} toward four more LMC stars confirmed the presence of outflowing gas and further found that the LMC's gas outflows correlated with H\textsc{~ii} regions and super shells, possibly signaling they are a result of supernovae in the disk. The blueshifted material, relative to the LMC, was also detected along sightlines that projected onto relatively quiescent regions. Lying in the direction of the LMC, \citet{Lehner2009} found further evidence that the high-velocity cloud (HVC) may have originated from an earlier LMC outflow event. They observed that this cloud has a velocity gradient with R.A. in \hi\ emission and low-ionization species (see their Figure~5), a similar oxygen abundance as the LMC, and depletion patterns that indicate the presence of dust ---a strong indicator that this material is of LMC origin.

However, \citet{Werner2015} were able to determine an upper distance limit of an HVC at a similar velocity that is positioned only a few degrees offset from the LMC's disk using absorption-line spectroscopy toward a halo star at a known distance. They found the cloud along $(l,~b)=(279\fdg9,~-37\fdg1)$ only lies $d_{\odot}\lesssim 13.3~\kpc$ away. While their distance limit provides compelling evidence that some of the HVC material at a local standard of rest (LSR) velocity of $\rm{v_{LSR}}=150~\kms$ is likely of MW origin \citep{Rich2015}, this does not eliminate the possibility of two separate HVC complexes. It remains that toward the LMC, an HVC is observed where the \hi\ position-velocity map shows a physical association with the LMC \citep{Staveley2003}, not the MW. It is one of the sole HVCs where dust depletion is observed \citep{Lehner2009}. The occurrence of both a MW and LMC HVC around the same projected area is plausible given the prior work and the large angular region being explored.

Each of the previously mentioned absorption-line studies that detected blueshifted material in the direction of the LMC use background targets embedded within the disk of the LMC, which only traces material between the MW and LMC. Moreover, the majority of previously explored sightlines were preferentially selected toward active star-formation regions where outflows are more likely to occur. However, there are four locations in the \citet{Lehner2007} that probe relatively quiescent regions that still find blueshifted material relative to the LMC, leaving open the possibility of a galactic wind across the whole of the near-side of the LMC disk. This comes in addition to the prospect of a hot corona coexisting with the wind \citep{deBoer1980,Wakker1998,lucchini2020}. The scenario of both an outflow and corona would support itself in that the outflow may feed the corona and supply it with gas.

In order to probe gas flows on the far-side of the LMC, \citet{Barger2016} used a ``down-the-barrel'' (star) and transverse (QSO) experiment to isolate material that is located on the far-side of the LMC disk. They found that both the low and high ions are symmetrically flowing out of the LMC disk at speeds representing an intermediate-velocity cloud (IVC). Those results, when combined with the previous studies of \citet{Howk2002} (absorption from 12~LMC stars showing outflowing material), \citet{Lehner2007} (additional 4 LMC stars correlating with H\textsc{~ii} bubbles and super shells), \citet{Lehner2009} (gradient found in the HVC toward the LMC), and \citet{Pathak2011} (strong absorption in OVI across the entire face of the LMC), provide convincing evidence that the LMC drives a global, large-scale outflow across its entire disk.

Our study supplements the previous UV work by supplying the first spectroscopically resolved \ha\ map of the LMC and its surrounding environment. This approach removes limitations of previous works that had small numbers of pointings that were dependent on the location of background targets, which severely reduced the spatial coverage of the material they observed. The work we present is unique in that (1) our observations are roughly an order of magnitude more sensitive than previous studies---enabling us to map the diffuse optical emission from these clouds---and that (2) we were able to spatially resolve the entirety of the LMC and its surroundings at an angular resolution of $\theta_{\rm resolution}=1\arcdeg$. Both of these results cannot be performed with the vast majority of other more distant gas-rich galaxies. With emission-line observations of the ionized component of the LMC's IVCs and HVCs, we explore their global morphology and kinematic distribution in Section~\ref{sec:channel}. In this section, we further assess whether the IVC material could be associated with an LMC wind origin. In Section~\ref{sec:ivc}, we discuss how a portion of the emission is from gas currently being driven from the galaxy as an IVC. This is followed up with an estimate of this material's mass as well as it's role in the LMC's neighborhood (Sections~\ref{sec:ivcmass} and \ref{sec:ivcdiscuss}). Section~\ref{sec:hvc} discusses a HVC that is moving at speeds upward of $\Delta {\rm v}_{\rm LMCSR}\approx -150~\kms$. The significance and possible explanation for this material's origin is considered in Section~\ref{sec:hvcdiscuss}.

\section{Data} \label{sec:data}
This study utilizes archival radio and newly acquired optical emission-line observations to trace the neutral and ionized hydrogen gas in and around the LMC.

\subsection{Wisconsin \ha\ Mapper}
We surveyed the faint \ha\ ($\lambda_{\rm H\alpha}=6562.8~\angstrom$) emission across the face of the LMC's disk and in the region that surrounds it using the Wisconsin \ha\ Mapper (WHAM) telescope over the $+50\le\rm v_{LSR}\le+250~\kms$ velocity range.\footnote{We use the kinematic definition of the LSR in which the Sun moves $20~\kms$ in the direction of $(\rm R.A.,~DEC.)=(18^h 3^m 50.29^s, ~30\arcdeg00\arcmin16\farcs8)$ for the Julian 2000 epoch (J2000).} Equipped with a Fabry-P\'erot spectrometer, WHAM is roughly an order of magnitude more sensitive than other currently available instruments, enabling us to detect the faint emission at a sensitivity limit of $I_{\rm H\alpha}\approx10~{\rm mR}$\footnote{A Rayleigh (R) is a unit of measure for the surface brightness of emission lines that is equal to $1~{\rm R}=10^6/{4\pi}~\rm photons\,cm^{-2}\,sr^{-1}\, s^{-1}$, which is $5.6\times10^{-18}~\rm erg\,s^{-1}\,cm^{-2}\,arcsec^{-2}$ for H$\alpha$.} per 30~second exposure. WHAM's sensitivity is achieved due to its high throughput resulting from its 1~degree beam size. By adjusting the gas pressures between the instruments etalon, we can tune our observations to center on the \ha\ emission-line that is associated with the LMC. More information about the WHAM telescope and its capabilities are outlined in \citet{Haffner2003}.

Our WHAM observations have a $\theta_{\rm resolution}=1\arcdeg$ angular resolution, which corresponds to a $\Delta d_{\rm resolution}\approx 1~\kpc$ spatial resolution at the distance of the LMC. The observations were Nyquist sampled across the face of the galaxy, effectively increasing the integration time per location on the sky and removing gaps between observations with overlapping pointings. Our \ha\ survey contains more than 6,600 individual observations that are positioned both on and off the LMC's \hi\ disk, covering an area that spans from $(l,b) = (246\fdg5,-17\fdg8)$ to $(315\fdg0,-46\fdg7)$.

\subsection{Radio Data}
We use archival \hi~21-cm emission-line data from the Parkes Galactic All-Sky Survey (GASS\footnote{The GASS survey is a publicly available survey with access through an online database retrieval site: https://www.atnf.csiro.au/research/GASS/Data.html.}; McClure-Griffiths+2009) to probe the neutral hydrogen gas phase in the LMC at an angular resolution of $\theta_{\rm resolution}=16\arcmin$, corresponding to an angular area that is roughly $14\times$ smaller than that of the WHAM beam. This survey spans a velocity range of $-400 \leq \rm v_{LSR} \leq +500~\kms$ and has a spectral resolution of $\Delta \rm v_{\rm resolution} =0.82~\kms$. The RMS brightness temperature noise is $T_{\rm B,~RMS}= 57~\rm mK$, corresponding to a $3\sigma$ sensitivity of $\log{\left(N_{\rm H\textsc{~i}}/\cm^{-2}\right)} \approx 18.2$ for a typical high-velocity cloud line width of $30~\kms$ \citep{mcclure}.\footnote{We convert \hi\ brightness temperatures ($T_{\rm B}$) to column densities using the relationship $N_{\rm H\textsc{~i}}= 1.823\times10^{18} \int ({T_{\rm B}}/\rm K)(d{\rm v}/\kms)~\cm^{-2}$, which assumes that these clouds optically thin $21~\cm$ radiation.} In this paper, the survey's $3\sigma$ limit is far lower than the practical limit used for our maps and mass calculation.

In addition to the GASS data, we use a combined \hi~21-cm emission-line map consisting of combined ATCA and Parkes telescope observations \citep{kim2003}. The velocity range of this LMC HI survey spans $+190\le\rm v_{LSR} \le +375~\kms$ and has a $\Delta\rm v_{\rm resolution}=1.6~\kms$ spectral resolution. Compared to the GASS survey, this combined map has a column density sensitivity of $\log{\left(N_{\rm H\textsc{~i}}/\cm^{-2}\right)} \approx 18.86$ for a $\Delta\rm v=30~\kms$ wide line, but a much higher spatial resolution at $\theta_{\rm resolution}=1\arcmin$. This data resolves smaller physical structures than GASS, improving our ability to discern smaller-scale morphological features in the disk.

We also use data from the Leiden/Argentine/Bonn (LAB) Survey to measure extinction along our sightlines due to its similar beam size as WHAM. LAB data has an effective FWHM beam of $\theta_{\rm resolution}=35\arcmin$ for declination $\leq -27\arcdeg$. The survey covers a velocity range of $-450 \leq \rm v_{LSR} \leq +400~\kms$ with a spectral resolution of $\Delta \rm v_{\rm resolution} =1.3~\kms$. The rms noise is $T_{\rm B,\,rms}=0.07~{\rm K}$ resulting in a $3\sigma$ sensitivity of $\log{\left(N_{\rm H\textsc{~i}}/\cm^{-2}\right)} \approx 18.38$.

\section{Data Reduction}
The H$\alpha$ reduction process we performed was carried out in two stages. In the first stage, we used the standard WHAM reduction pipeline, which performs the bias subtraction, flat-fielding, ring-summing, cosmic ray contamination removal, and air mass corrections. In the second stage, we velocity calibrated our spectra, removed atmospheric signatures from observations, masked out observations affected by foreground stars, and corrected for dust extinction.

\subsection{WHAM Pipeline}
We utilized the WHAM pipeline that is described in detail in \citet{Haffner2003}. During this data processing, pixels warmed by cosmic rays were first removed. The circular interference patterns that result from our Fabry-P\'erot spectrometer observations were summed in annuli to produce a linear spectrum that is a function of velocity. These linear spectra span a $\Delta\rm v = 200~\kms$ velocity range and are uniformly binned to $\Delta\rm v_{\rm bin}=2~\kms$ intervals. The pipeline normalizes the spectra by exposure time, scales them for the air mass of observations, and applies an intensity correction factor to account for sensitivity degradation of the WHAM instrumentation that occurs over time.

\subsection{Velocity Calibration}
Our observations span $+50 \leq \rm v_{GEO} \leq +250~\kms$ in the geocentric (GEO) velocity frame. Over this range, these observations do not overlap with the bright geocoronal \ha\ line at $\rm v_{GEO}=-2.3~\kms$ and only overlaps with the blue wing of a bright OH line at $\rm v_{GEO}=+272.44~\kms$. Therefore, we were unable to use either of these lines to calibrate our velocities using the method described in \citet{hausen2002} and \citet{Barger_2013}. Instead, we used the velocity calibration technique that is described by \citet{Barger2017} and \citet{antwi-danso2020} for WHAM observations that do not overlap with bright atmospheric lines at well established transitions. Using this technique, we calibrated our velocity by monitoring the pressure of the $\rm SF_6$ gas in the WHAM Fabry-P\'erot etalons and by further refining the calibration by comparing our observations with an atmospheric template.

Using the linear relationship between the pressure of the Fabry-P\'erot etalons and $\Delta \lambda$ measured by \citet{tufte1997}, we calculated the velocity offset between the raw and geocentric velocity frames. This is essentially the reverse of our tuning process, which enabled us to calibrate the velocity frame to an accuracy of $\Delta\rm v_{GEO}\lesssim5~\kms$. Because all of our observations were taken at the same tune (i.e., at the same interference order), the relative velocities of the calibrated observations agree within $0.1~\kms$ of each other as described in \citet{Barger2017}. We improved our calibration further by aligning our observations with the faint atmospheric lines in the atmospheric template presented by \citet{Barger_2013} (see their Figure~3). This enabled our velocity solution to be calibrated to an accuracy of $\Delta\rm v_{GEO}\lesssim1~\kms$. We then converted from GEO to LSR using a constant offset that accounted for the date, time, and location of each observation.

\subsection{Atmospheric Subtraction}
\subsubsection{Atmospheric Template}\label{sec:atm_temp}
There are significant faint atmospheric emission features which populate the entire velocity range of our observations. While these lines are abundant, they behave predictably and vary primarily with air mass. Because of this, we modelled these lines using a template created by \citet{Barger_2013} (see their Figure~3), which characterizes the atmospheric emission present in our observations. The template was scaled to account for differences in air mass between observations and night-to-night variations due to humidity and temperature.

Brighter lines are more variable and need to be fit individually. This includes a bright OH molecular line at $\rm v_{GEO}=272.44~\kms$ whose line strength depends on the angle between the Sun and Earth's upper atmosphere. In the direction of the LMC, this bright OH line is overwhelmed by emission from the LMC's disk. To subtract this emission feature in the direction of the LMC, we kept the area and width of the line fit constant so that it matched with off-disk observations that were taken during the same night and within roughly 15 minutes of the on-disk observations. Because this narrow OH line is kinematically unresolved by the WHAM telescope, we fit the line assuming that it has a width of $\Delta \rm v_{\rm OH~width}=1~\kms$ before we convolved it with WHAM's $\Delta \rm v_{\rm WHAM~IP} \approx 12~\kms$ instrument profile.

\begin{figure*}[t]
\centering
\includegraphics[angle=90,width=0.75\textwidth]{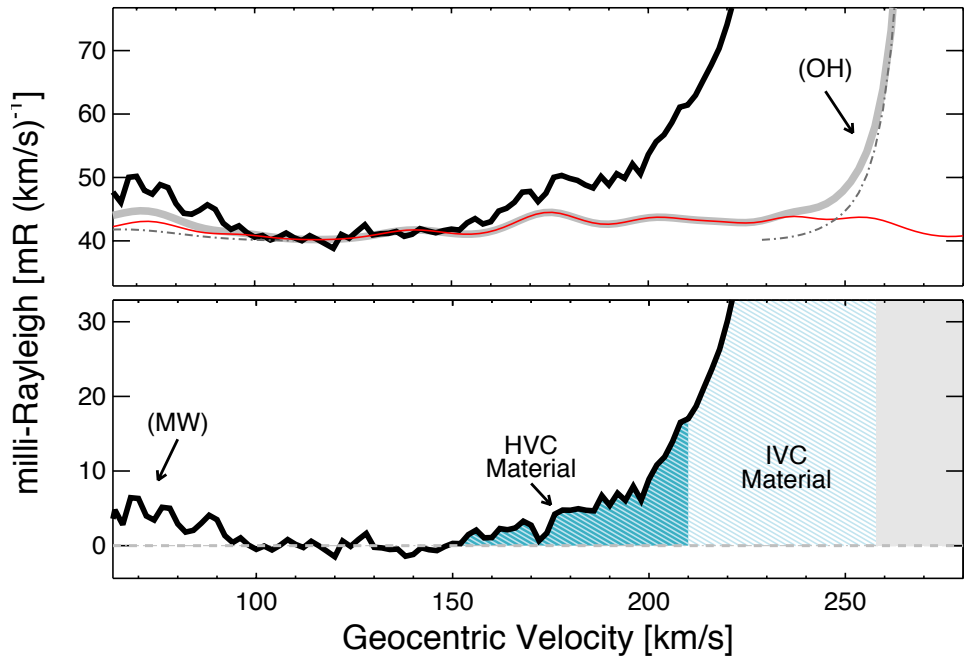}
\caption{(Top panel) A pre-atmospheric subtracted WHAM spectrum toward $(l,~b)=(279\fdg0,~-31\fdg0)$ drawn as the black line. The atmospheric template we used to reduce our \ha\ observations is indicated with a red overlaid line. The emission contributions associated with the Milky Way at $\rm v_{GEO}=+66.21~\kms$ and a bright OH line at a velocity of $\rm v_{GEO}=+272.44~\kms$ are traced with dashed dot gray lines. The solid, thick gray line traces all of the atmospheric emission (atmospheric template and OH line) that we subtracted from the WHAM spectrum during our reduction procedure. (Bottom panel) The final reduced spectrum, where emission from the LMC ($\rm v_{GEO}\gtrsim+260~\kms$) is shaded light gray, LMC IVC material ($+210\lesssim \rm v_{GEO} \lesssim+260~\kms$) is shaded light blue, LMC HVC material ($+150\lesssim \rm v_{GEO} \lesssim+210~\kms$) is shaded dark blue, and the Milky Way ($\rm v_{GEO}\lesssim+90~\kms$) is marked.}
\label{fig:initial_reduction_spec}
\end{figure*}

We fit the background emission with a constant term added to the atmospheric template. The total fit ---which includes the bright OH line at ${\rm v_{geo}}=272.44~\kms$, faint atmospheric lines, a flat background, and Gaussian modelled astronomical emission--- utilized a chi-square minimization with conservative criteria to avoid over-fitting emission features that are not physically realistic. These criteria account for the line width and lower limits for the strength of the astronomical emission, which are dependent on the gas temperature and instrument sensitivity, respectively. An example of the pre and post atmospheric corrected spectrum is shown in Figure~\ref{fig:initial_reduction_spec} with a sightline piercing through the LMC. A detailed description of these bright lines and how they were handled, including their origin and the nature of their variability, is outlined in \citet{Barger_2013}.

\subsubsection{Removal of Systematic Signatures}

\begin{figure}[ht]
\centering
\plotone{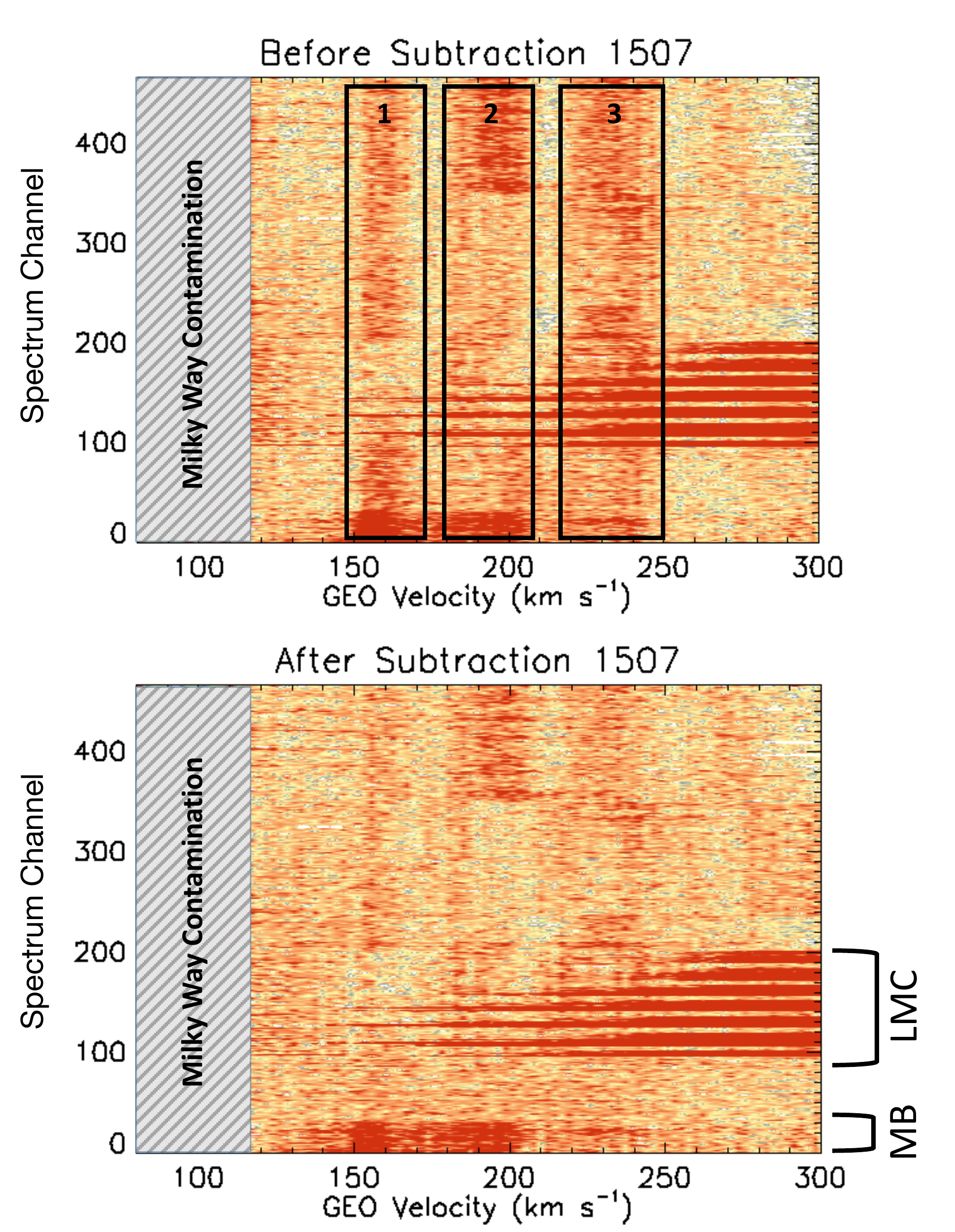}
\caption{\replaced{(Top) A latitude sub-region of 466 WHAM spectra before the correction of atmospheric}{(Top) A latitude sub-region between $-45\le b \le -36$ containing 466 WHAM spectra before the removal of residual} signatures. At various locations of $\rm v_{GEO}=+155$, $+195$, and $+230~\kms$, vertical structures are visible within the blue dashed rectangles, labeled 1, 2, and 3. (Bottom) After the correction there is a large improvement in the removal of the vertical signatures. At those same velocities as the top panel, the signatures are reduced. In both panels, emission from the LMC is visible as horizontal stripes over channels 100--200 while the Magellanic Bridge (MB) is visible from channels 0--35 at lower velocities.}
\label{fig:geo_loops_final}
\end{figure}

Following the removal of the atmospheric contamination with the atmospheric template described above, we discovered systematic spectral signatures in the reduced spectra. The cause of these structured residuals is likely due to very faint, unresolved atmospheric lines that are not described in the atmospheric template or from a slight velocity misalignment between the observed spectra and the atmospheric template causing the signatures during subtraction. These signatures appear at the same geocentric rest frame (GEO) velocity over narrow, $5\lesssim\Delta\rm v\lesssim10~\kms$, velocity widths. These signatures are visible when the spectra in our survey are stacked across the same velocity range in the geocentric frame, as shown in the top panel of Figure~\ref{fig:geo_loops_final}. At several velocities in this frame, there are multiple relatively coherent vertical signatures across the spectra. However, the two bright astronomical horizontal structures are associated with the Magellanic Bridge ($+150\lesssim\rm v_{GEO}\lesssim+210~\kms$ and $0\le\textrm{spectrum channel}\le35$) and the LMC and its wind ($+150\lesssim\rm v_{GEO}\lesssim+300~\kms$ and $100\le \textrm{spectrum channel}\le200$).

These faint residual atmospheric signatures exist across all observed spectra. To characterize these spectral artifacts, we averaged the spectra together at locations within our map that contain little to no astronomical \ha\ emission above our sensitivity limit. We then subtract this average off target spectrum from our observed spectra to effectively remove these signatures. To ensure that the off target spectra accurately characterized the sky for that region of the map, this procedure was repeated for four Galactic latitudinal sub-regions. We display stacked spectral image with before (top panel) and after (bottom panel) samples of this reduction process for a subset of our observations in Figure~\ref{fig:geo_loops_final}.

Overall, the vertical atmospheric features at $\rm v_{GEO}\approx+150$, $+180$, and $+230~\kms$ have been greatly reduced. However, some of this residual atmospheric emission remains, especially at $\rm v_{GEO}\approx+180~\kms$ for $\textrm{spectrum channels}>350$ that correspond to a region on the sky between $(l,\,b)=(265\arcdeg,\,-20\arcdeg)$ and $(285\arcdeg,\,-25\arcdeg)$. These atmospheric residuals persist in these spectra because the Galactic latitude region had few good \ha\ faint off locations. Although the presence of this lingering atmospheric emission in our final reduced \ha\ emission map is not ideal, it does not impact the final results of this paper as we do not use the data in that region of the sky in our mass calculation.

\begin{figure}[ht]
\centering
\plotone{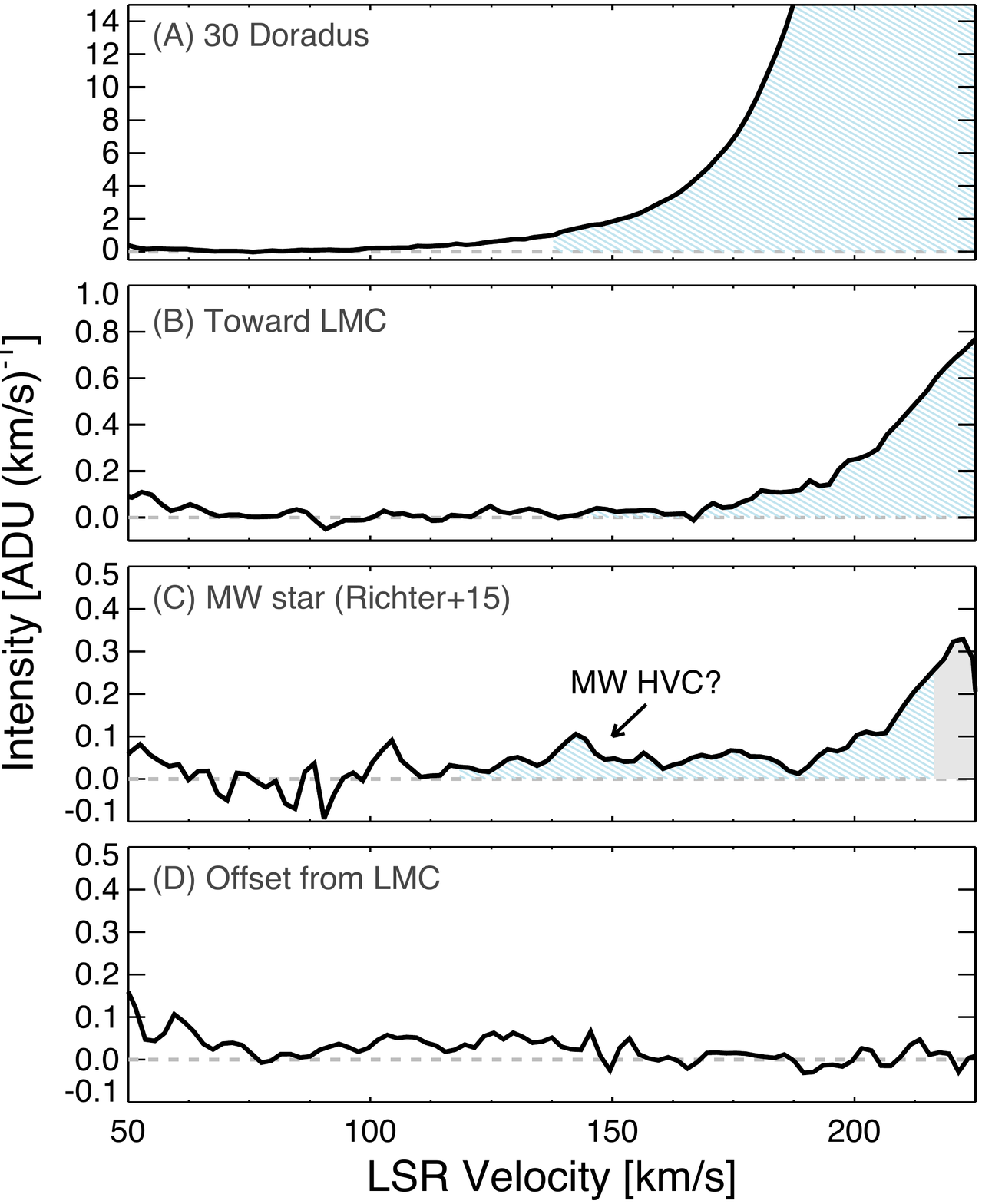}
\caption{Four example \ha\ spectra that probe different positions in our survey. This includes emission toward: (Panel~A) 30~Doradus at $(l,~b)=(279\fdg5,~-31\fdg7)$, (Panel~B) the LMC's disk at $(278\fdg3,~-31\fdg0)$, (Panel~C) the MW star used in the \citet{Rich2015} study $(279\fdg9,~-37\fdg1)$, and (Panel~D) off the LMC's disk at $(271\fdg7,~-29\fdg7)$. In these panels, we shade material that coincides with the \hi\ disk of the LMC in light gray (only Panel~C over the displayed LSR velocities) and LMC IVC material out to a line-of-sight velocity of $|\rm  v_{LOS}|=100~\kms$ of its disk in light blue.}
\label{fig:spectra}
\end{figure}

\subsection{Observational Cutoffs and Degradation Correction}
We removed sightlines that are within a $0\fdg55$ angular radius of stars that are $m_{\rm V} \le 6.0~{\rm mag}$ as their stellar continuum contributes non-linear emission to the \ha\ spectra. This cutoff removes 689~observations from our sample, corresponding to roughly 12\% of our observations. Because WHAM is a remote observatory configured for queue observing and does not require constant monitoring during good weather conditions, we double checked all of the observations that were taken at the optimal CCD temperature for the WHAM camera of $T_{\rm CCD}=-101.2\arcdeg{\rm C}$ and not during an automated liquid nitrogen fill to reduce noise. We confirmed that the etalon pressures were stable and that the monitored values matched the input values during setup for the observations at night. We also ensured that all of the used observations were taken when the outside humidity was less than 85\%, at a Zenith Distance less than $75\arcdeg$, and during dark time observations. Throughout this process a total of 291 additional observations (or around 4\%) of the remaining sample were removed. In total, there were 980~observations (or 14\%) removed from our sample leaving 5,931 observations. For sightlines with repeat observations, we averaged the spectra together. Our survey samples a total of \sightlines unique sightlines, where each sightline was observed an average of 3.5~times with the locations toward the LMC's \hi\ disk sampled the most. 

As a result of WHAM instrument degradation over time, observations suffered up to a 20\% decrease in observed \ha\ intensity \citep{smart2019}. The procedure used for determining the WHAM instrument degradation trend with time is outlined in \citet{Haffner2003}. However, our intensity correction does not include a night-to-night intensity calibration associated with airmass variations that are due to variations in atmospheric conditions. This is because there were insufficient calibration observations taken each night during this survey, in part because there are few WHAM calibration targets in the southern sky.

\subsection{Extinction Correction}
Previous absorption-line spectroscopic studies toward LMC stars (e.g., \citealt{Howk2002,Lehner2007,Lehner2009,Barger2017}) indicate that the gas clouds surveyed in this study exist between us and the LMC. Additionally, \citet{Barger2017} found compelling evidence that the gas within $\Delta \rm v_{LOS}\approx100~\kms$ from the \hi\ disk of the LMC along the line of sight is associated with large-scale galactic outflow. Similarly, the depletion patterns observed by \citet{Lehner2009} for the HVC material that is in the same projected region on the sky also indicates that this cloud contains dust. We, therefore, applied two extinction corrections: one for attenuation due to the Milky Way and another for self-extinction of the gas clouds assuming that they have a chemical composition similar to that of the LMC. 

To correct for reddening, we used the following relationship from \citet{Diplas1994} that relates the color excess with the average $N_{\rm H\textsc{~i}}$ foreground emission:
\begin{equation}
    E(B-V)=\frac{\left<N_{\rm H\textsc{~i}}\right>}{4.93\times10^{21}~\rm atoms/(cm^2~mag)}
\end{equation}
We used the Leiden/Argentine/Bonn Galactic \hi\ survey (LAB; \citealp{KalberlaLAB,Hartmann}) to calculate the average foreground \hi\ emission. For the MW, we integrated the \hi\ emission over the $-100\le \rm v_{LSR}\le+100~\kms$ velocity range and assumed an extinction parameter of $R_{\rm v}=3.1$. Similar to the \citet{Barger_2013} study, we also adopt the \citet{Cardelli1989} optical extinction curve. Combined, the total extinction for the MW dust is then given by:
\begin{equation}
    A(\rm H\alpha)=5.14\times10^{-22}\,\left<N_{\rm H\textsc{~i}}\right>\,\rm cm^{-2}\,atoms^{-1}~mag,
\end{equation}
where the extinction corrected intensity is then $I_{\rm H\alpha,\,corr}=I_{\rm H\alpha,\,obs} e^{A(\rm H\alpha)}$. After correcting for extinction associated with foreground MW material, our observed \ha\ intensities increase by roughly 10\%.

The self extinction by the circumgalactic medium is small as this gas is diffuse. Following a similar process to the MW extinction, we can account for the self extinction of the winds by adopting the extinction parameter for the LMC measured by \citet{Gordon2003} of $R_{\rm v}=3.41$. We used an $R_{\rm v}$ that is measured for the LMC's disk as the IVC and HVC material likely originated from this galaxy via stellar feedback events. When we integrated the \hi\ emission across the velocity range of our wind, $+100\le \rm v_{LSR}\le+225~\kms$, we found that the associated self extinction correction is much less than 1\% and, therefore, we neglected this correction in our mass calculations below.

\section{LMCSR Velocity Frame}
\begin{figure}
\centering
\plotone{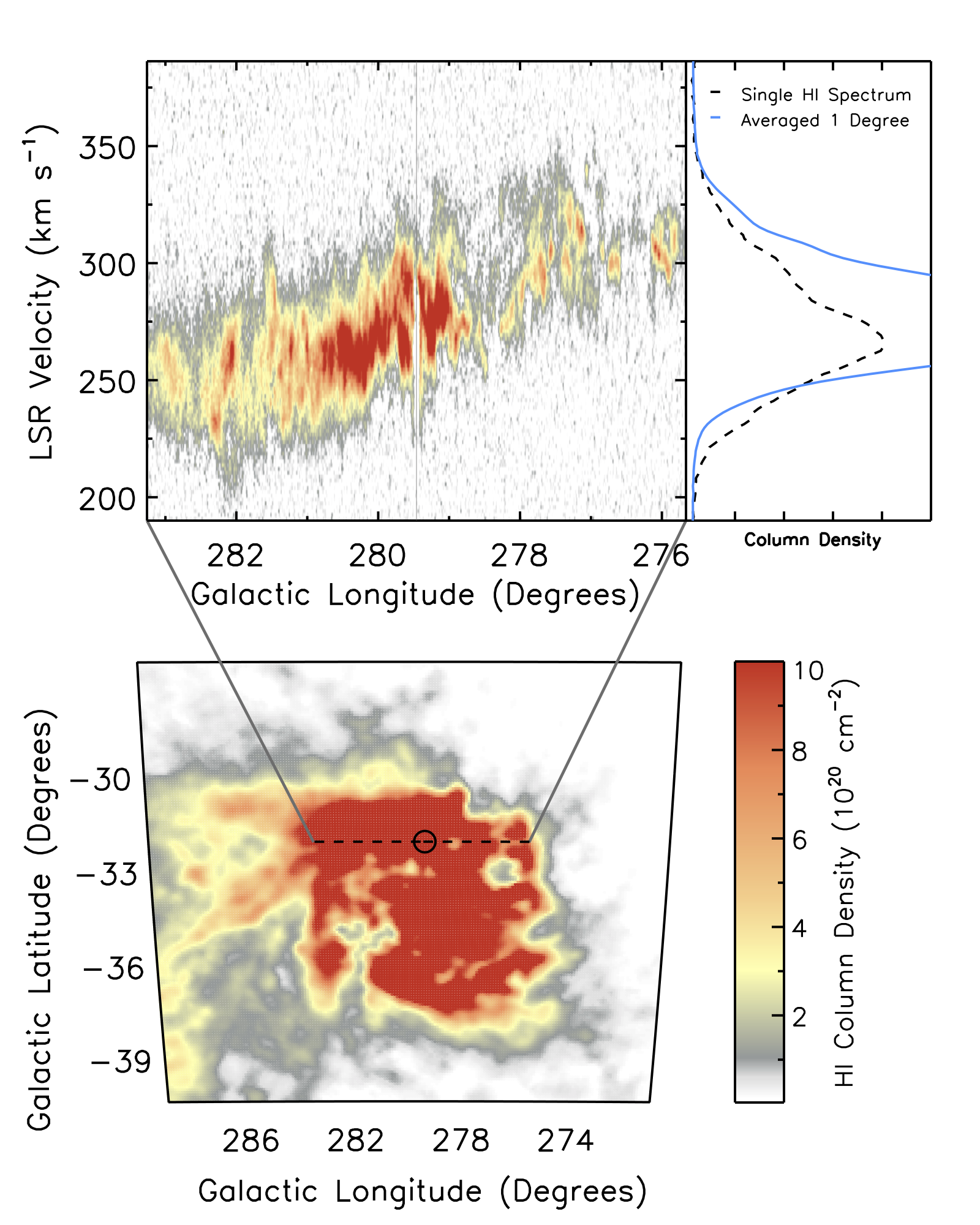}
\caption{(Bottom) An \hi\ intensity-weighted position-position map of the LMC. The circle marks the location of 30~Doradus at $(l,~b)=(279\fdg46,~-31\fdg67)$ and the dashed line indicates the extent of the Galactic Longitude range considered the top-left panel. (Top-left) A position-velocity map of \hi\ emission running through the location of 30~Doradus. (Top-right) \hi\ spectra toward the 30~Doradus sightline. The dashed line represents a single ATCA \hi\ spectrum toward 30~Doradus while the solid blue curve depicts an average spectrum for all emission within circular area of 1~degree diameter centered on 30~Doradus.}
\label{fig:posvel}
\end{figure}

When exploring the circumgalactic material of the LMC, it is useful to use a reference frame centered around the LMC disk rather than the LSR frame. We refer to this reference frame as the Large Magellanic Cloud Standard of Rest (LMCSR) frame. Because the LMC is actively forming stars across its \hi disk, the kinematic width of its disk varies rapidly at small scales. Across a slice through the LMC's disk at Galactic latitude of $b=-31\fdg67$ and centered on the 30~Doradus starburst region, the width varies from $25\lesssim\Delta {\rm v}_{\rm H\textsc{~i}}\lesssim50~\kms$ (see Figure~\ref{fig:posvel}). This multiple component structure complicates the process of determining where the disk kinematically ends and where a wind begins. 

Across the roughly 10~degree Galactic longitude slice shown in Figure~\ref{fig:posvel}, the motion of the \hi\ gas has a velocity gradient that spans from $+225\lesssim {\rm v_{LSR}} \lesssim+275~\kms$ at higher Galactic longitudes of $l\approx282\arcdeg$ to $+275\lesssim {\rm v_{LSR}} \lesssim+325~\kms$ at $l\approx276\arcdeg$. Along this velocity slice, there are several locations containing holes where little to no \hi\ exists. This is not surprising as these holes can be created by energetic stellar feedback process activity occurring inside the disk, which heats and ionizes the surrounding gas. This feedback can further drive circumstellar and interstellar material outward, possibly contributing to a galactic outflow as noted by \citet{Staveley2003}. In the LMCSR velocity frame, the spatially varying velocity gradient in the LSR frame is removed, which helped us disentangle the disk and the wind material.

To convert our spectroscopic observations from the LSR to LMCSR velocity reference frame, we initially used the relationship provided by \citet{Lehner2009}, which described the motion of the LMC's disk stars. However, galaxy interactions have disrupted the LMC's disk so that the gaseous and stellar components do not align. While the \citet{Lehner2009} LSR to LMCSR relationship works reasonably well for the motion of the gas toward the disk, our observations extend $\Delta\theta\gtrsim5\arcdeg$ off the LMC's gaseous disk on all sides and is no longer centered in that velocity reference frame. Instead, we modeled the \hi\ emission the LMC and its surroundings to convert our \ha\ observations into a LMCSR velocity reference frame. This is especially beneficial because the gaseous \hi\ emission extends much further than the stellar disk and because \ha\ emission tends to kinematically follow the \hi\ emission in HVCs (e.g., \citealt{Haffner2001,Putman2003,Hill2009,barger2012,Barger2013,Barger2017,antwi-danso2020}).

We determined the motion of the LMC's \hi\ disk by performing a Gaussian decomposition of \hi\ GASS spectra across our surveyed region. We enforce that the Gaussian fits meet the following criteria: each fit must have a column density above $\log{\left(N_{\rm H\textsc{~i}}/\cm^{-2}\right)} \approx 19$, kinematic width of approximately $\rm 30~\kms$, and a velocity centroid between $+175\leq {\rm v_{LSR}} \leq+325~\kms$ to be considered part of the LMC disk. We modelled the \hi\ disk as a simple 2D plane using a least-squares fit. To improve the accuracy of this plane, we weighted our fit by the \hi\ column density. Our resultant relationship between the line-of-sight Galactic longitude ($l$) and latitude ($b$) and the central velocity offset is:
\replaced{
\begin{equation}\label{eqn:vlmcsr}
\frac{\Delta \rm v_{LMCSR}}{\kms}=1293.22 -3.25\left(\frac{l}{\deg}\right) + 3.66 \left(\frac{b}{\deg} \right)
\end{equation}
}{
\begin{equation}\label{eqn:vlmcsr}
\frac{\Delta \rm v_{LMCSR}}{\kms}=262.55 -3.25\left(l-280\right) + 3.66 \left(b-33 \right)
\end{equation}
}
This offset corresponds to an LMCSR velocity as follows: ${\rm v_{LMCSR}}$=${\rm v_{LSR}}$+${\Delta \rm v_{LMCSR}}$. We used the width of the \hi\ lines out to $3$~standard deviations to describe the thickness of the LMC's \hi\ disk and to distinguish between disk and wind material. The near-side and far-side disk boundaries are described by the difference between the $3$~standard deviation fitted plane and the central velocity (Equation~\ref{eqn:vlmcsr}). This results in adopting an LMC disk width of roughly $80~\kms$ across the face of the LMC, or $40~\kms$ from the kinematic center of the disk to its edge. The newly constructed velocity frame improves our ability to separate the wind material from the LMC's disk that lies in front of the galaxy and to identify the wind material that extends past its H\textsc{~i} disk.

\section{Kinematic Morphology of \ha\ Emission}\label{sec:channel}

\begin{figure*}[t]
\centering
\plotone{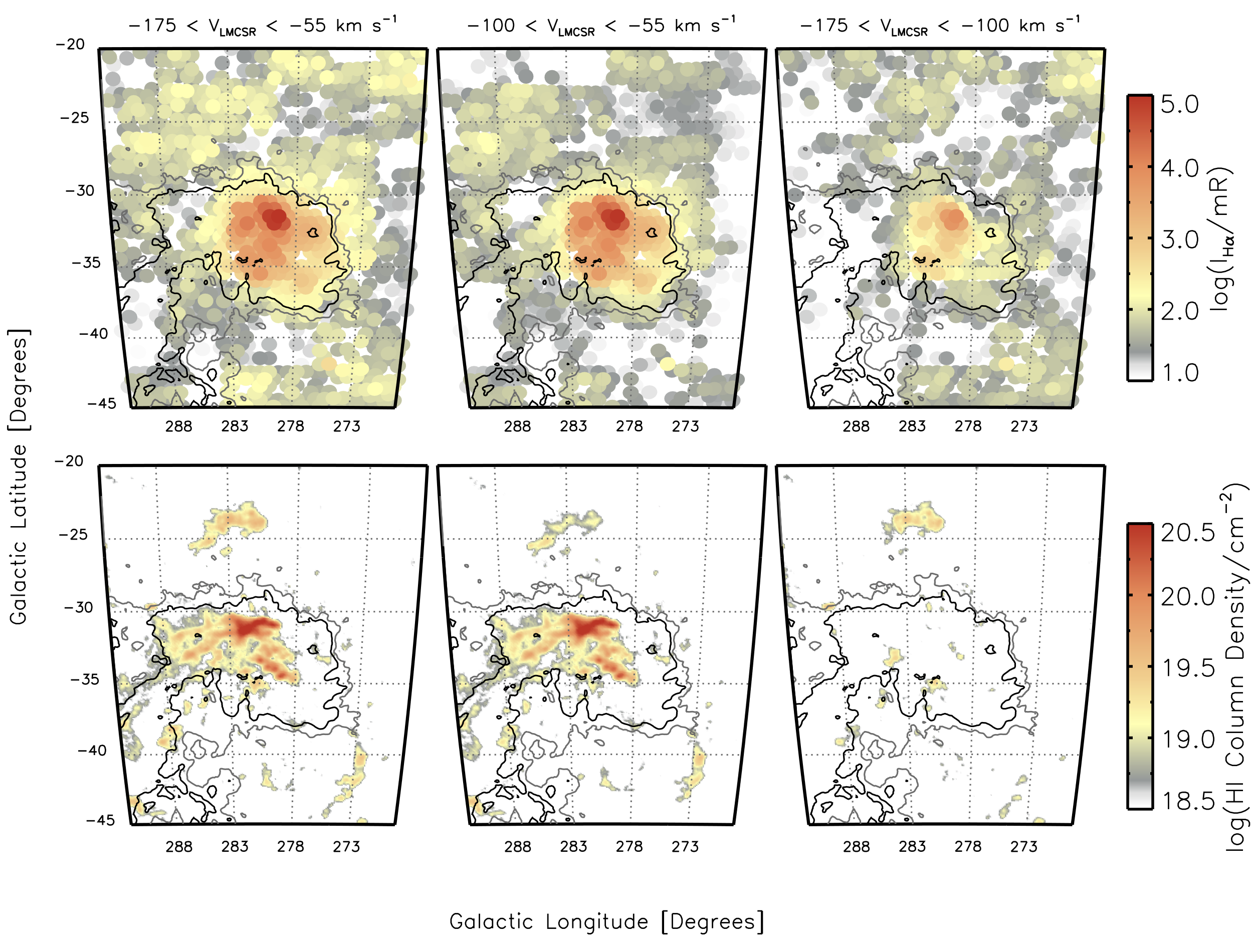}
\caption{(Top) \ha\ emission maps of the LMC and its surroundings. This map traces the material that is blueshifted relative to the LMC. From left to right, the total wind integrated over the $-175 \leq \rm v_{LMCSR} \leq -55~\kms$ velocity range, the IVC portion integrated over the $-100 \leq \rm v_{LMCSR} \leq -55~\kms$ velocity range, and the HVC portion integrated over the $-175 \leq \rm v_{LMCSR} \leq -100~\kms$ velocity range. The overlaid black contours trace the \hi\ emission across the same integration range at $\log(N_{\rm H\textsc{~i}}/{\rm cm}^{-2})=19$. (Bottom) \hi\ column density map covering the same region and velocity ranges above using GASS data.}
\label{fig:EM_total}
\end{figure*}

We observe blueshifted material at the intermediate- and high-velocities relative to the LMC in both \hi\ and \ha\ emission. This emission is consistent with the results of previous UV absorption-line spectroscopy studies that suggest a large-scale galactic wind emanating from the LMC, driven by the stellar activity within its disk \citep{Howk2002,Lehner2007,Barger2016}. The \hi\ IVC emission is strong toward the LMC's disk and rapidly decreases radially. The \ha\ emission similarly decreases radially away from the LMC yet extends off the boundary of the LMC stellar ($r_\star\sim2.15~\kpc$) and \hi\ disk at $\log{\left(\rm H\textsc{~i}/\cm^{-2}\right)}\approx19$ (see the left-hand panel of Figure~\ref{fig:EM_total}). This \ha\ emission is asymmetric, relative to the H\textsc{~i}, such that it extends farther along the edge of the LMC disk near the 30~Doradus starburst region.

Gaseous debris has littered the surrounding area of the LMC due to its interactions with the SMC. Therefore, in addition to the wind that is likely associated with the LMC, there is Magellanic tidal material and MW HVCs that pollute this region of the sky. At higher Galactic latitudes than the LMC ($b\geq-27\arcdeg$), there are a few sparse \hi\ clouds that are likely associated with the Leading Arm (LA) complex LA~I near $(l,~b)\approx(283\arcdeg,\,-24\arcdeg)$; for more details on the \ha\ distribution of these offset clouds, see \cite{Smart2021}, in preparation. Likewise, because the Magellanic Bridge connects the LMC and SMC, its emission is present at $l\geq285\arcdeg$ with similar velocities as the high Galactic longitude edge of the LMC. Toward the southern rim of the LMC disk, there are fragmented clouds centered on $(l,~b)=(273\arcdeg,~-41\arcdeg)$, $(278\arcdeg,~-38\arcdeg)$ and $(281\arcdeg,~-42\arcdeg)$ (see the right-hand panel of Figure~\ref{fig:EM_total}). The background halo star that \citet{Rich2015} used to establish the distance of an HVC ($d_\odot\leq13.3\kpc$) at the location $(l,\,b)=(279\fdg9,\,-37\fdg1)$ lies within the region of these low-latitude fragmented clouds. These clouds overlap in velocity with the HVC absorption (${\rm v}_{\rm LMCSR}\approx-150~\kms$) along this stellar sightline, suggesting that it could be associated with the MW. For this reason, we conservatively do not consider gas that is more than a few degrees off of the LMC's \hi\ disk to be part of the LMC outflow.

\begin{figure*}[t!]
\centering
\includegraphics[width=\textwidth]{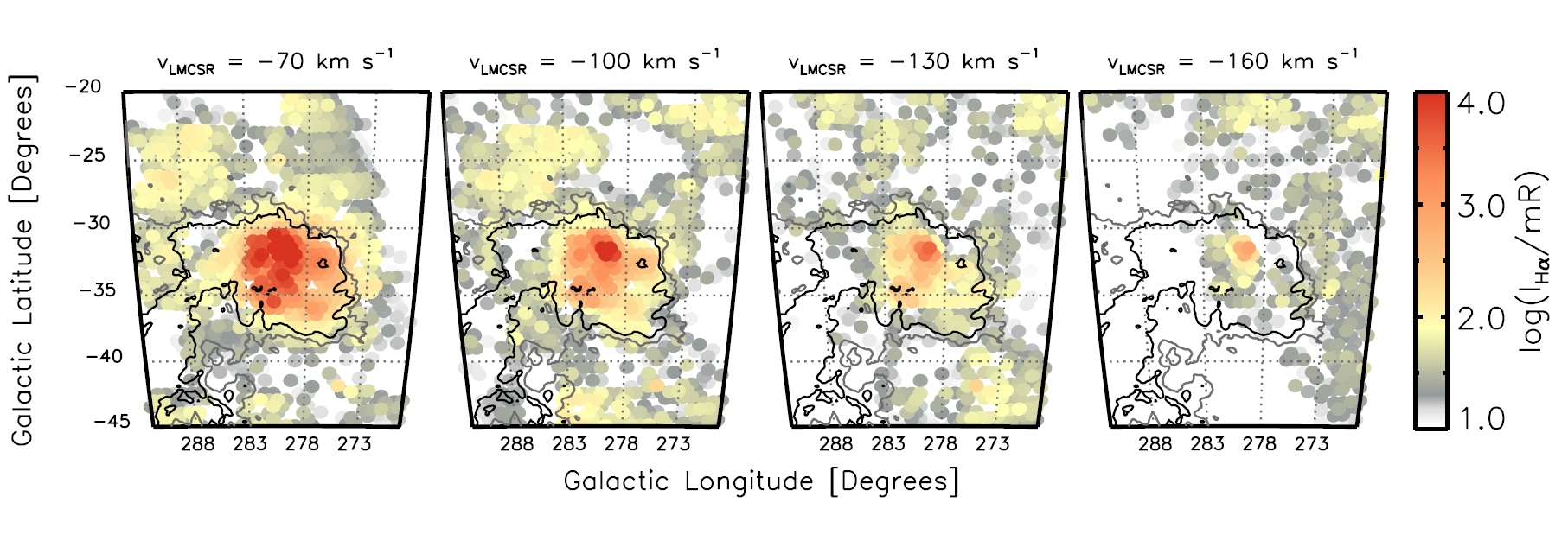}\\
\caption{\ha\ emission channel maps centered at $\rm v_{LMCSR}=-70$, $-100$, $-130$, $-160~\kms$ (left to right). Widths of these velocity slices are all $\Delta{\rm v}=30~\kms$. The bright emission in the right-hand panel at ${\rm log} (I_{\rm H\alpha}/{\rm mR})\approx2.5$ spatially aligns with the 30~Doradus starburst at $(l,~b)=(279\fdg5,~-31\fdg7)$. \hi\ column density contours are drawn at $\log(N_{\rm H\textsc{~i}}/{\rm cm}^{-2})=19.0$ and $20.0$ in gray and black, respectively.}
\label{fig:CM_ha}
\end{figure*}

We find the gaseous material toward the LMC that is the least blueshifted (i.e., kinematically closest to the LMC's motion) morphologically follows the \hi\ disk of the LMC. In Figure~\ref{fig:CM_ha}, we separate the \ha\ into four separate emission maps, each with small integration ranges that allow us to study the bulk properties of the gas cloud in discrete slices of velocity. The emission at ${\rm v}_{\rm LMCSR}\approx-100~\kms$ maintains an intensity well over $I_{\rm H\alpha}\approx0.3~\rm R$ across the face of the LMC (See left two panels of Figure~\ref{fig:CM_ha}). This widespread \ha\ emission of the IVC is consistent with a large-scale LMC outflow.

At higher velocities, approaching the more blueshifted material, our data shows the emission has a strong spatial alignment with the most active star-forming region within the LMC, 30~Doradus (see right two panels of Figure~\ref{fig:CM_ha}). This emission spans over $\Delta {\rm v}_{\rm LOS}\gtrsim150~\kms$ and remains stable across each channel map toward the star-forming region. Detecting strong \ha\ emission out to velocities of nearly $\Delta {\rm v}_{\rm LMCSR}\approx -150~\kms$ while also observing a connection to the lower velocities in the same region indicates an association with the LMC.

\section{Intermediate Velocity Gas}\label{sec:ivc}
We find numerous clouds that are bright in \ha\ emission that are blue shifted by roughly $50-100~\kms$ relative to the LMC's \hi\ disk (Figures \ref{fig:EM_total} and \ref{fig:CM_ha}). Most of this intermediate velocity material spatially aligns with the disk of the LMC, resembling a galactic outflow previously suggested \citep{Howk2002,Barger2016}. Using emission across \sightlines \ha\ sightlines toward the LMC IVC, we determine its mass, its mass-flow rate, and mass-loading factor. In the following LMC IVC mass calculations, we only include the material that is within $\Delta\theta\approx1\arcdeg$ of the LMC \hi\ disk with $\log{\left(\rm H\textsc{~i}/\cm^{-2}\right)}\gtrsim19$. This is because some of the \ha\ emission that is projected multiple degrees off its disk could be associated with Magellanic tidal debris (i.e., Magellanic Bridge, Leading Arm) or MW HVCs (see Figure \ref{fig:EM_total} and Section \ref{sec:channel}).

\subsection{Mass Estimate of IVC}\label{sec:ivcmass}
\citet{Barger2016} estimated the mass of the intermediate velocity outflowing LMC winds to be $\log{\left(M_{\rm ionized}/M_\odot\right)}\gtrsim7.16$ for the low-ionization species. Because they found that this wind is roughly symmetrical on either side of the LMC's disk, this would correspond to a mass of $\log{\left(M_{\rm ionized}/M_\odot\right)}\gtrsim6.86$ for only the near-side ionized outflow. However, as that study only sampled the wind along two neighboring sightlines via absorption-line spectroscopy, they had to make assumptions about the spatial extent and morphology of the wind. With our kinematically resolved, near-side \ha\ emission map of the wind of the LMC, we are able to measure both of these directly.

In contrast with the absorption-line work, our \ha\ survey allows us to obtain a mass estimate more naturally from the wind's density times its volume, $M=\rho {\rm V}$. We calculate its mass density using the electron number density as a proxy for the density of protons as they are roughly equal (i.e., $n_p\approx n_e$) and use a reduced mass of $\mu\approx1.4 m_{\rm H}$ to account for the contribution from helium and metals. Calculating the volume of the wind requires knowing its solid angle $\Omega$, line-of-sight depth $L$, three~dimensional geometry (see Figure~\ref{fig:volumes}), and distance $D$. We also include a $\cos{\left(i\right)}$ factor to account for the inclination of the cross-sectional area of the wind relative to our line of sight. The mass is then given as: $M=\mu n_e \Omega D^2 L\cos{\left(i\right)}$. For gas at the distance of the LMC, the mass enclosed within one WHAM beam with $\Omega=1\arcdeg$ is then:
\begin{equation}\label{eqn:mass}
    \frac{M_{\rm ionized}}{{M_\odot}} = 2.1~\times~10^4 \cos{\left(i\right)} \Bigg(\frac{D}{50~{\rm kpc}}\Bigg)^2 \Bigg(\frac{L_{{\rm H}^+}}{\rm pc} \Bigg)
    \Bigg(\frac{n_e}{\rm cm^{-3}}\Bigg) 
\end{equation}

We estimate the total mass of the wind by summing the single beam mass across the projected outflow area. For the \sightlines \ha\ spectra that fill our map of the LMC (Figure~\ref{fig:EM_total}), we define the morphological extent of the LMC's galactic wind to include regions that are within $1\arcdeg$ of its \hi\ disk with neutral column densities larger than $\log(N_{\rm H\textsc{~i}}/\cm^{-2})\geq19.0$. This region contains 215 WHAM sightlines contained within roughly $50~{\rm deg}^2$ that are used for our mass estimate. For each of these sightlines, we integrated across the $\rm -100\le v_{LMCSR} \le-55~\kms$ velocity range to measure the \ha\ intensity of this wind and to explore its spatial distribution (see Equation~\ref{eqn:vlmcsr}). The strength of the \ha\ recombination line is directly proportional to the electron density squared along the line-of-sight depth and the electron temperature ($T_e$) of the gas as:
\begin{equation}\label{eqn:intensity}
\frac{I_{\rm H\alpha}}{\rm R}=0.364\, T_4^{-0.924} \left(\frac{EM}{{\rm pc}\,\cm^{-6}}\right),
\end{equation}
where EM is the emission measure ($EM\equiv\int n_e(s)^2 ds$) and $T_4$ is given in units of ten thousand Kelvin $(\rm i.e., T_4=T_e/{10^4\,K})$. Measurements of the average EM and associated velocity range are given in Figure~\ref{tab:velEM}.

Since we cannot measure the line-of-sight depth or the electron density as a function of depth directly, we adopt a few necessary assumptions to estimate the mass of the wind that closely follow the procedures used in prior WHAM \ha\ studies for evaluating the mass of HVCs (e.g., \citealt{Hill2009,barger2012, Barger2017,smart2019}).

\begin{deluxetable}{ccc}
\tablecolumns{3}
\tablecaption{Observed Velocities and Emission Measures\label{tab:velEM}}
\tablehead{
\colhead{Outflow Component}  & \colhead{$v_{\rm LMCSR}$} & \colhead{$\left<{\rm EM}\right>$\tablenotemark{a}} \\
\colhead{}\vspace{-0.5cm} & \colhead{$(\kms)$} & \colhead{$(\rm 10^{-3} pc\,cm^{-6})$} \\
}
\startdata
IVC	&	$-100$ to $-55$		&	$390$	\\
HVC	&	$-175$ to $-100$	&	$205$	\\
\enddata
\tablenotetext{a}{This is an average emission measure across the corresponding velocity range used to calculate the ionized mass across all sightlines considered to be part of the wind.}
\end{deluxetable}

\subsubsection{Line-of-Sight Depth}
The most difficult of these assumptions pertains to the depth and line-of-sight distribution of the wind. Past studies of the HVC component of the LMC wind observed similar kinematics for the neutral and low-ionization species (e.g., \citealt{Lehner2009}). Moreover, observations of both \ha\ and \hi\ in outflows of other galaxies (e.g. M82; \citealt{Lehnert1999} and \citealt{Schwartz2004}) support a multi-phase wind that is well mixed at large scales. In our study, due to our large angular resolution at 1~degree, we are spatially resolving the wind at the kiloparsec scale and are unable to resolve small-scale structure in the wind. We therefore assume that the neutral gas and ionized gas are well mixed at the scales we are probing in our survey such that the ionized hydrogen depth is roughly equal to the neutral depth, i.e., $L_{\rm H^+}\approx L_{\rm H\textsc{~i}}$. 

\begin{figure}[t!]
\plotone{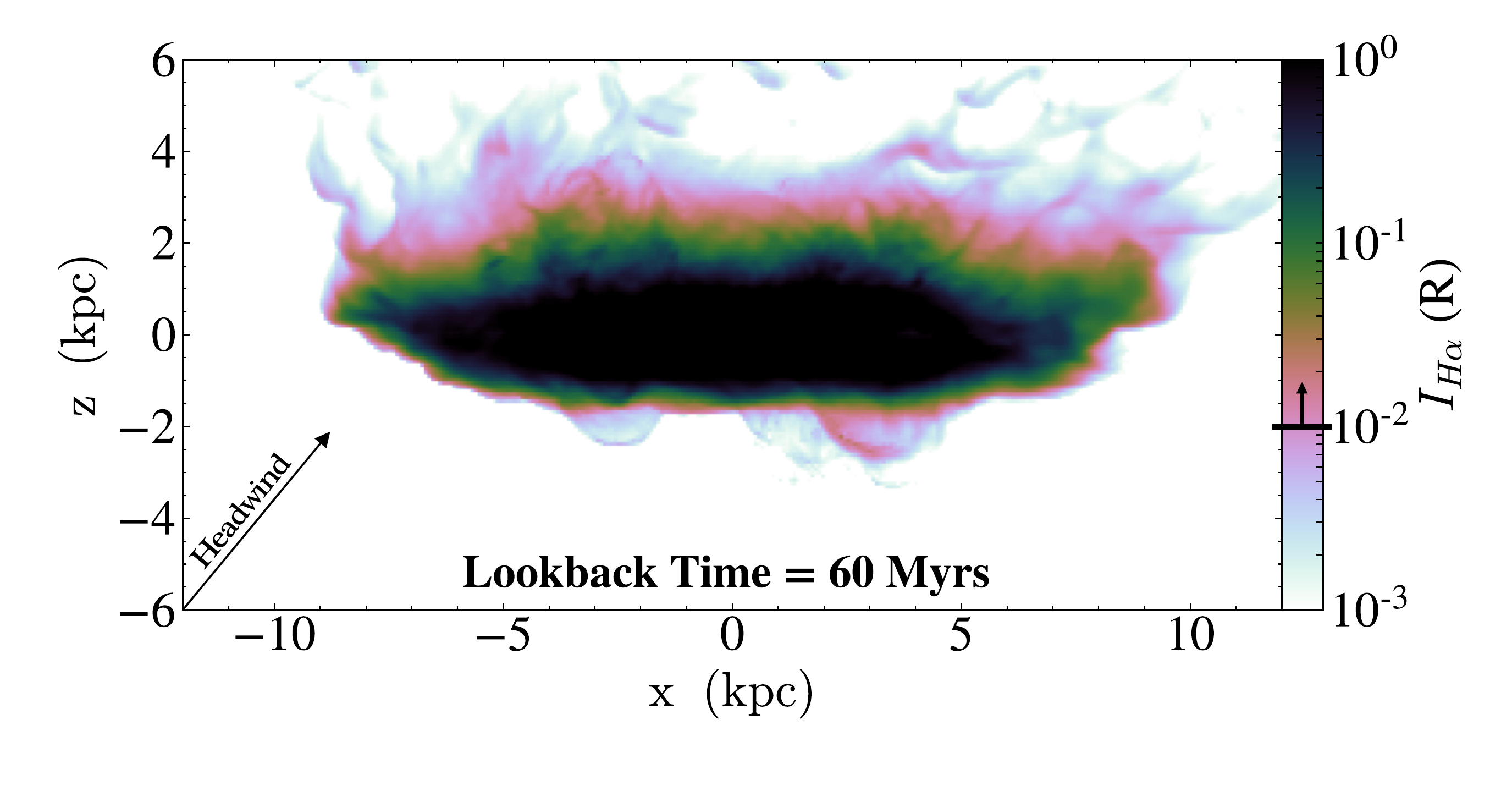}
\plotone{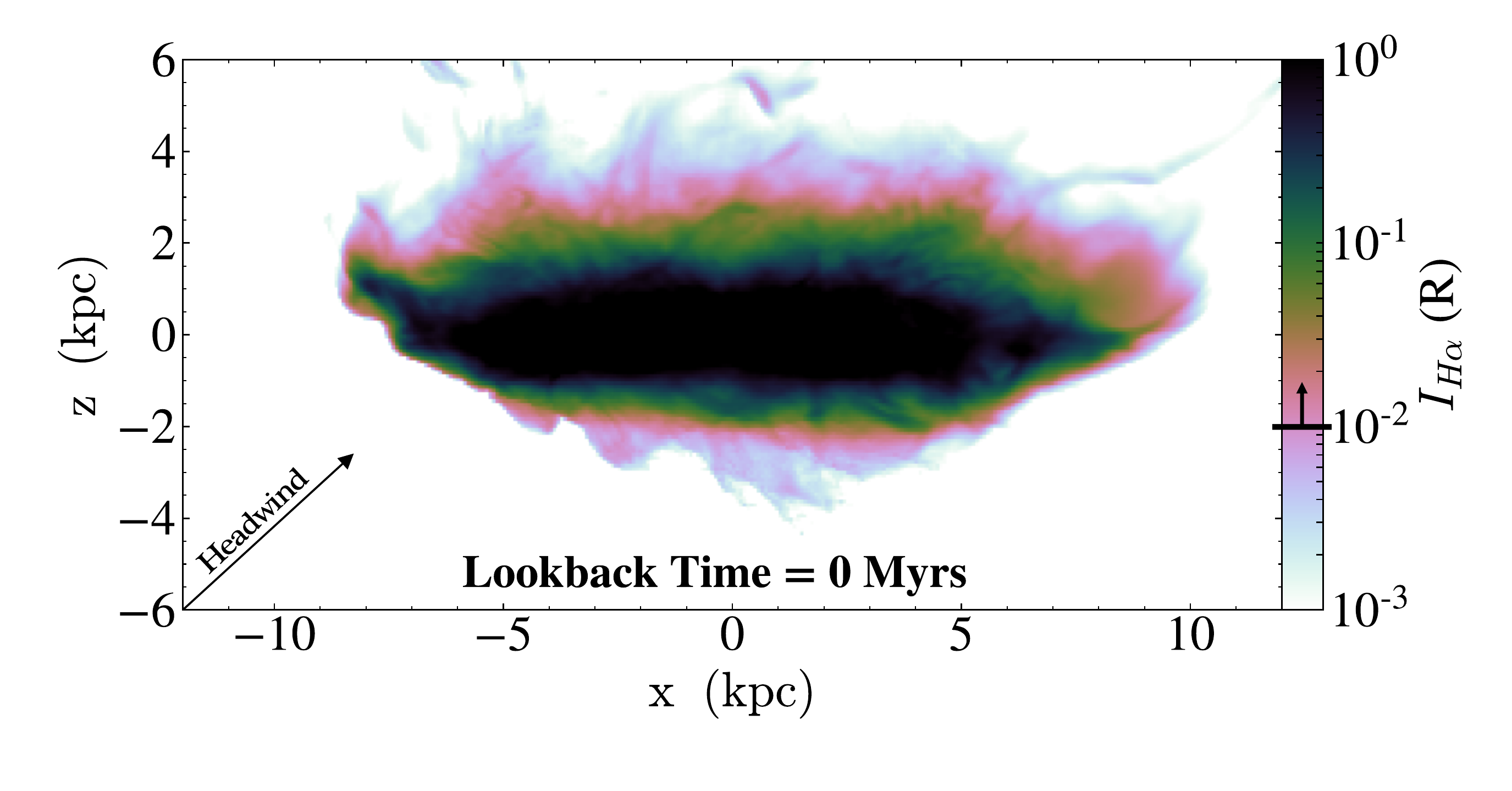}
\caption{Simulated \ha\ emission maps from \citet{bustard2020} of the LMC's galactic wind from an edge-on perspective, using the Trident package \citep{hummels2017}. This wind is assumed to be in photoionization equilibrium with the UV background (no local ionizing sources from the LMC or Milky Way are included). This model uses the orbital history of the LMC from the models of \citet{Besla2012}, the present-day infall velocity of the LMC is $258~\kms$ directed edge-on and $194~\kms$ directed face-on with respect to the LMC; the ambient medium is assumed to be smooth with a total gas number density of $\sim 10^{-4}~{\rm cm}^{-3}$ at the present-day LMC distance of $d_\odot\approx 50~{\rm kpc}$ \citep{salem2015}. (Top) \ha\ emission at a lookback time of $60~{\rm Myr}$. (Bottom) Current day edge-on view of the LMC and its \ha\ emission. The arrow in the lower-left corner of each panel represents the motions of the head wind caused by the LMC's path through the MW halo. On each colorbar, the sensitivity of our observations (10~mR) is marked with a horizontal line and an arrow pointed upward.}
\label{fig:chad}
\end{figure}

To estimate this depth, we analyze the fiducial simulations of \citet{bustard2020} for LMC-specific outflows (see Figure~\ref{fig:chad}). These magnetohydrodynamic simulations used the observed LMC star-formation history from \citet{Harris2009} to the seed star cluster particles that would subsequently deposit the thermal, kinetic, and cosmic ray energy into surrounding grid cells. In these simulations, gas that emerged from the LMC's disk due to stellar driven outflows experienced an external pressure by surrounding coronal gas. \citet{bustard2020} found that the apparent coronal gas wind pushes against the leading edge of the LMC via ram pressure and that its effects are strong enough to suppress the near-side outflows and alter the shape of the LMC’s halo. While this simulation neglects the gravitational influence of the SMC and Milky Way, we expect the depth of the outflow to be primarily influenced by the LMC’s gravitational potential and ram-pressure effects.

On a global scale, the galactic winds produced in the \citet{bustard2020} simulations match well kinematically and spatially with the observed LMC outflow. We use the \citet{bustard2020} results as a guide for constraining the depth of this wind. They find that the $10^4~{\rm K}$ gas in the near-side outflow penetrated to a height of $z_{\rm wind}\approx 3~{\rm kpc}$ below the midplane of the LMC ($z_{\rm midplane}=0~{\rm kpc}$) at a lookback time of $60~{\rm Myrs}$. At present-day, they find that this wind stalls at a height of $z_{\rm wind}\approx 4~{\rm kpc}$ due to ram pressure. The outflows on the far-side of the LMC, however, are able to travel much further off the disk as ram-pressure effects are weaker on the trailing side of galaxy. Accounting for the height of the LMC's \hi\ disk ($z_{\rm H\textsc{~i}\,disk}\approx1.75~{\rm kpc}$), this corresponds to present-day wind depth of roughly $1 \leq L_{\rm wind} \leq 2~\kpc$ off the galaxy.
%\citep{subramanian2009}

We used this depth to calculate an average electron density for the LMC's galactic wind using the measured EM as \replaced{$\langle n_e\rangle=EM^{1/2} L_{\rm H^+}^{-1/2}$}{\textbf{$\langle n_e\rangle=\langle EM\rangle^{1/2} L_{\rm H^+}^{-1/2}$}}. We then used this density to calculate the outflow's mass with Equation~\ref{eqn:mass}. Although our main uncertainty involves the depth of the wind, it is important to note that the mass scales as $M_{\rm ionized}\propto L_{\rm H^+}^{-1/2}$, resulting in only a modest variance in mass when we consider a range of reasonable depths. Moreover, we assume an electron temperature in the range $0.8\lesssim T_4\lesssim 1.2$, which is where the \ha\ emission peaks. We further assume that the temperature of the neutral and ionized hydrogen gas are roughly equal allowing us to relate the neutral and ionized hydrogen number densities for a given pressure scenario as $P_{\rm ionized}/n_{\rm ionized}=P_{\rm neutral}/n_{\rm neutral}$ under ideal gas conditions.

\begin{figure}[t!]
\plotone{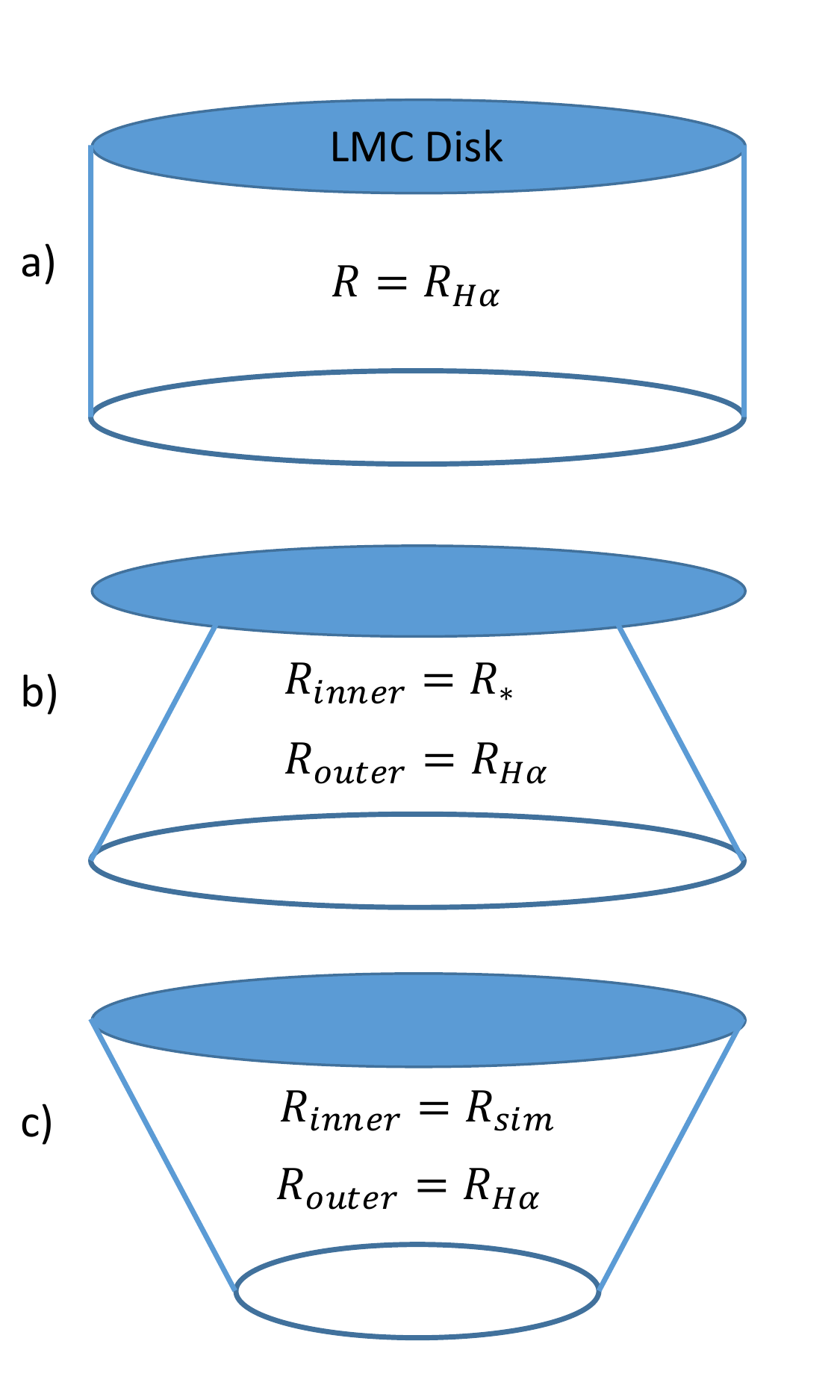}
\caption{Explored near-side volume geometries of the LMC outflow, including (a) a cylinder with a uniform outflow that spans the face of the galaxy out to some height, (b) a partial cone with its narrow end embedded and centered on a region of intense star-formation, (c) an inverted partial cone with its narrow side pointing away from the LMC's disk to match the morphology of the simulated galactic outflow under influence of ram-pressure and headwinds.}
\label{fig:volumes}
\end{figure}

Because the LMC is nearly face-on, we cannot constrain how the morphology of the wind varies with depth as we only see its 2~dimensional projection on the sky. Therefore, we acknowledge three separate volume scenarios: cylinder, partial outward flaring cone, and partial tapered cone (see Figure~\ref{fig:volumes}). For the cylindrical wind in scenario (a), we simply assume that the radius of the wind is constant and matches the extent of the \ha\ emission. We include the outward flaring partial cone geometry in scenario (b) as it has been observed for other galaxies (e.g., M82); in this scenario, we set the inner cone radius to the stellar radius (2.15~kpc) and the outer radius to match radius of the \ha\ emission. The tapered, inverted cone in scenario (c) is the geometry that resulted for the near-side LMC wind from \citet{bustard2020} simulation when they accounted for ram-pressure effects; in this scenario, we match the radii to match the simulation and the \ha\ observations. For these three geometries, the near-side wind masses would correspond to $\log{\left(M_{\rm ion}/M_\odot\right)}=7.36\pm0.14$ for volume~(a),  $\log{\left(M_{\rm ion}/M_\odot\right)}=7.10\pm0.14$ for volume~(b), a mass of $\log{\left(M_{\rm ion}/M_\odot\right)}\leq7.09\pm0.14$ for volume (c). As we cannot observationally determine which of these wind geometries better matches with the LMC's near-side galactic wind, we will report the values of the cylindrical scenario in the text as it is the simplest volume that requires the least assumptions. We report the values and ranges for the other two volume scenarios in Table~\ref{tab:mass_results}. To estimate the total mass of the neutral and ionized gas of this wind, we assume the outflow is symmetric on both the near-side and far-side of the LMC's disk and that it has an ionization fraction of ${ n_{\rm H^+}/n_{\rm H}\approx0.75}$ \citep{Barger2016}. We find the total IVC mass of the wind to be in the range $7.70\leq\log{\left(M_{\rm total}/M_\odot\right)}\leq7.85$. This is compared to the previous \citet{Barger2016} estimate of $\log{\left(M_{\rm ionized}/M_\odot\right)}\gtrsim7.16$.

We calculate the mass-flow rate for this wind by assuming an outflow time ($t_{\rm outflow}\approx60~\rm Myr$). This is calculated using information regarding the last period of star-formation ($\sim100~{\rm Myrs}$) as well as the time necessary for the wind to penetrate through the surrounding medium and travel approximately 2~kpc off the \hi\ disk. This results in a total IVC mass-flow rate of $0.83\leq\dot{M}_{\rm outflow}\leq1.18~M_\odot~\rm yr^{-1}$. The mass-loading factor is also calculated to study the ratio of the mass-flow rate to the star-formation rate, $(\eta\equiv \dot{M}_{\rm outflow}/\dot{M}_\star)$. We adopt a star-formation rate in the range $0.3\lesssim\dot{M}_\star\lesssim0.34~M_\odot~\rm yr^{-1}$ to agree with the star-formation history of the LMC (see Figure~11 of \citealt{Harris2009}). This results in a mass-loading factor between $2.44\leq\eta\leq3.93$. Because the mass-loading factor, $(\eta\equiv \dot{M}_{\rm outflow}/\dot{M}_\star)$, is much greater than unity, this indicates that the current star-formation state of the LMC is unsustainable such that the galaxy could become quenched if this state is prolonged and if the ejected gas is able to escape.

\subsection{IVC Material - Discussion}\label{sec:ivcdiscuss}

\citet{Barger2016} characterized the IVC material with respect to the LMC using UV absorption-line spectroscopy toward a LMC disk star and a background QSO. They found the near-side material to have an estimated mass of $\log{\left(M_{\rm low~ions}/M_\odot\right)}\gtrsim7.16$ for low-ionization species on both the near-side and far-side of the galaxy. This corresponds to an ionized mass of $\log{\left(M_{\rm ionized}/M_\odot\right)}\gtrsim6.9$ for gas traveling up to $\Delta {\rm v}_{\rm LMCSR}\approx -100~\kms$ on the near-side of the LMC. Over the same velocity range, we calculated the ionized hydrogen mass to be $\log{\left(M_{\rm ionized}/M_\odot\right)}\approx7.36$ (see Section~\ref{sec:ivc}). While our estimate is larger than the \citet{Barger2016} near-side mass, it is important to note the discrepancies between our estimates can be attributed to how each study determined the masses.

In the \citet{Barger2016} study, they assumed (1) the wind has an angular extent similar to the LMC's \hi\ disk ($R_{\rm H\textsc{~i}}\approx3.7~\kpc$), (2) a covering fraction of $f_{\Omega}=0.7$ for low-ionization and $f_{\Omega}=0.9$ for high-ionization species, and (3) the average global strength for this wind could be represented by the absorption they observed along their two sightlines. We, however, find that the \ha\ emission of the wind extends around 1 degree on average off the \hi\ disk, which means that the outflow radius is roughly $1~\kpc$ larger (i.e., $R_{\rm H\alpha}\approx R_{\rm H\textsc{~i}}+1~\kpc$). Their third assumption may result in a significantly underestimated outflow mass as the two sightlines used in the \citet{Barger2016} study probed a relatively quiescent region of the LMC. We emphasized the effect of their assumptions by taking the average emission measure from the same region as the \citet{Barger2016} sightline---in the opposite quadrant as 30~Doradus---and calculating a corresponding mass for an area similar to theirs. Using $\langle EM\rangle\approx0.2~{\rm pc}\,{\cm^{-6}}$, the outflow mass would be $\log{\left(M_{\rm ionized}/M_\odot\right)}\approx7.1$, which is nearly half as large as our IVC cylindrical mass estimate and nearly in agreement with the \citet{Barger2016} estimate for the low-ionization species.

Ultimately, the fate of this ejected material remains uncertain. The escape velocity of the LMC is roughly $110~\kms$ \citep{Besla2015b}, which corresponds to roughly $100~\kms$ along the line of sight for an inclination of around 25-degrees. Therefore, the majority of the IVC material is not expected to escape. However, the Magellanic System is a crowded environment and tidal interactions between the SMC, and possibly the MW, can assist in the removal of otherwise bound gas (see \citealt{Donghia2016} for a review). Some of the material from previous outflow events may have been displaced into the trailing Magellanic Stream. In a kinematic investigation of the \hi-21cm emission of the Magellanic Stream, \citet{nidever2008} found that one of its two filaments traces back to the 30~Doradus region of the LMC. \citet{richter2013} measured the chemical composition of this ``30-Doradus'' filament and found that it has a metallicity that is consistent with an LMC origin. \citet{fox2013} explored the chemical composition of the other Magellanic Stream filament and found that it has a lower metallicity that is more consistent with an SMC origin, which kinematically traces back to the Magellanic Bridge \citep{nidever2008}.

Ram-pressure stripping may also play an important role removing gas from the Magellanic Clouds. The impact that ram pressure has on the CGM strongly depends on the density of the medium that it is traveling through and on the motion of the gas within its surrounding medium. Based off the work of \citealt{salem2015}, \citealt{bustard2020} and others (\citealt{Heckman2000}, \citealt{Mastropietro2005}, and \citealt{fujita2009}), the effects of ram pressure on gas is multi-faceted and either can promote or suppress the removal of gas from a galaxy depending on the circumstance. This is because ram pressure can work in direct opposition of galactic winds positioned on the leading side of a galaxy, where the surrounding coronal gas will act as a head wind that pushes against the outflowing back toward the galaxy. Therefore, the outflow on the near-side of the LMC galaxy, which is the side leading the LMC's orbit through the MW's halo, will be suppressed. Meanwhile, the far-side could experience an enhanced outflow as ram pressure will push the gas away from the galaxy and it could therefore be more massive. With ram-pressure stripping, the simulated wind in the \citet{bustard2020} study was able to reach a height of more than $1.7~\kpc$ off the LMC's \hi\ disk and had a total ejected mass in the range $6.7\leq\log{\left(M_{\rm ejected}/M_\odot\right)}\leq8.4$. While our IVC mass estimate is within this mass range, \citeauthor{bustard2020} predicts that a significant portion of this material flows off the far-side of the LMC's disk (see their Figure~11), which is the trailing side trails as the LMC traverses the MW's halo. Conversely, the \citet{Barger2016} UV absorption-line study found that the near-side and far-side outflows were kinematically symmetric along their explored sightlines, though this might not be representative of the outflow's large-scale structure. A study like this one of the far-side---in which the winds are mapped---is needed to determine its physical extent and the impact of ram-pressure stripping.

\section{HVC Material}\label{sec:hvc}
We also detected a high-velocity component to the LMC's galactic wind in \ha\ emission over the $-175\leq\rm v_{LMCSR}\leq-100~\kms$ velocity range (see top-right panel of Figure~\ref{fig:EM_total}). Much of this material is traveling away from the LMC at speeds that exceed its escape velocity and could therefore be permanently lost from the galaxy. Furthermore, tidal forces could assist in carrying this material away. At these high velocities, the \ha\ emission is especially concentrated in the direction of the 30~Doradus starburst region (refer to the right panel of Figure~\ref{fig:CM_ha}). We calculate the ionized mass of this HVC using the procedures described in Section~\ref{sec:ivcmass}.

Using the timescale of the IVC material (60~Myrs) as well as the observed velocities, we estimate the HVC reaches a height up to 9~kpc off the \hi\ disk. It is possible the material reaches even further as \citet{bustard2020} shows ejected material to reach heights in excess of 13~kpc. With these estimates and equation~(\ref{eqn:mass}), we calculate an ionized hydrogen mass of $\log{\left(M_{\rm ionized}/M_\odot\right)}=6.99\pm0.13$ for the HVC. This mass contains 124 WHAM beams that are within roughly $20~{\rm deg}^2$. Following the procedure in Section~\ref{sec:ivcmass}, we assume the wind to be symmetrical about the LMC disk and to contain both neutral and ionized material. This corresponds to a total mass of $\log{\left(M_{\rm total}/M_\odot\right)}\approx7.41$, but we emphasize that this could be a lower limit on its total mass as ram-pressure effects are likely enhancing the far-side wind (see Figure~\ref{fig:chad} and \citealt{bustard2020}). Given that this material is the HVC component of the current outflow explored in Section~\ref{sec:ivcmass}, the mass-flow rate is $\dot{M}_{\rm outflow}\approx0.43~M_\odot~\rm yr^{-1}$ and the mass-loading factor is $\eta\approx1.36$. The full ranges of these results are provided in Table~\ref{tab:mass_results}.

When comparing our mass to prior work there is a general agreement. \citet{Lehner2009} estimated its neutral hydrogen gas mass to be $5.7\leq\log{\left(M_{\rm H\textsc{~i}}/M_\odot\right)}\leq6.0$ with a total mass of at least $\log{\left(M_{\rm total}/M_\odot\right)}>6.7$ with the assumptions that it has an LMC origin and that it lies at a distance of $40\leq d_\odot\leq50~\kpc$ away. Using the ionization fraction of $0.5\lesssim n_{\rm H\textsc{~II}}/n_{\rm H}\lesssim0.8$ that \citet{Lehner2009} measured, we calculate a total hydrogen mass (neutral and ionized) in the range $6.9\leq\log{\left(M_{\rm H}/M_\odot\right)}\leq7.1$, which is in agreement with their total mass lower limit. Moreover, simulations from \citet{bustard2020} estimated $6.56\leq\log{\left(M_{\rm ejected}/M_\odot\right)}\leq7.78$ worth of material reaching over $13\kpc$ away from the LMC disk. If we consider that the bulk of the HVC material distance of the wind to be at $d_\odot\approx13~\kpc$, then our observed HVC mass falls within the range of this simulated ejected mass.

\begin{deluxetable*}{ccccc}
\tablecolumns{5}
\tablecaption{Mass Estimates for IVC \& HVC Material of the LMC outflow \label{tab:mass_results}}
\tablehead{
\colhead{Outflow Geometry}\vspace{-0.2cm}  & \colhead{$M_{\rm ion}$} & \colhead{$M_{\rm total}$}  & \colhead{$\dot{M}_{\rm outflow}$} & \colhead{$\eta$} \\
\colhead{}\vspace{-0.5cm}  & \colhead{$(M_\odot)$} & \colhead{$(M_\odot)$}  & \colhead{$(M_\odot~\rm yr^{-1})$} &  \\
}
\startdata
\textbf{IVC}	&		&		&  &		\\\hline
Cylinder	&	$7.28$ -- $7.43$	&	$7.70$ -- $7.85$ & $0.83$ -- $1.18$ &	$2.44$ -- $3.93$	\\
Outward Cone	&	$7.02$ -- $7.17$	&	$7.44$ -- $7.59$ & $0.46$ -- $0.65$ &	$1.35$ - $2.16$	\\
Inward Cone	&	$7.01$ -- $7.16$	&	$7.43$ -- $7.58$ & $0.45$ -- $0.63$ &	$1.32$ -- $2.11$	\\\hline
\textbf{HVC}	&		&		&  &		\\\hline
Cylinder	&	$6.93$ -- $7.05$	& $7.35$ -- $7.47$	& $0.37$ -- $0.49$ & $1.09$ -- $1.63$	\\
\enddata
\tablecomments{All values for $M_{\rm total}$ assume a symmetrical wind on the near-side and far-side of the LMC with an ionization fraction of $n_{\rm H\textsc{~ii}}/n_{\rm H}=0.75$ that includes neutral and ionized gas assuming a reduced mass of $\mu\approx1.4 m_{\rm H}$ to account for helium and metals. We used the values listed in Table~\ref{tab:velEM} to calculate these outflow masses and rates.}
\end{deluxetable*}

\subsection{Origins of the HVC}\label{sec:hvcdiscuss}

We observe high-velocity material toward the LMC at velocities greater than $100~\kms$ off the LMC \hi\ disk ($\rm +90\le v_{LSR} \le+175~\kms$), detailed in Section~\ref{sec:hvc}. This emission is persistent at intermediate- to low-velocities relative to the LMC and spatially aligns well with the LMC's \hi\ disk (see Figure~\ref{fig:CM_ha}). These observed properties are consistent with an LMC origin in which the gas is expelled from the galaxy by its stellar activity. This is a conclusion that has also been reached by \citet{Staveley2003} using \hi\ emission-line and by \citet{Lehner2009} using UV absorption lines (also see \citealt{Lehner2007} and \citealt{Barger2016}). \citet{Staveley2003} found that the \hi\ column densities of the HVC peaked at locations that align with \hi\ voids within the LMC disk (such as supergiant shells, e.g., LMC~3); they further identified spatial and kinematic \hi\ bridges that linked back to the LMC's disk. \citet{Lehner2009} used 139~stars embedded in the LMC as background targets in an UV absorption-line study to explore the properties of this HVC; they found that the HVC has (i) a LSR velocity gradient in right ascension that follows the LMC's velocity gradient, (ii) dust based on depletion patterns---signifying a galactic origin (see also \citealt{Smoker2015}), (iii) an oxygen abundance similar to the LMC of $[\rm O\textsc{~i}/H\textsc{~i}]=-0.51^{+0.12}_{-0.16}$, and (iv) a high covering fraction in the direction of the LMC.

However, since the works mentioned above, the origin of this HVC has been strongly debated. This is because \citet{Werner2015} found C\textsc{~ii}, Si\textsc{~ii}, and Si\textsc{~iii} absorption consistent with this HVC in the spectra of a background star at a distance of $d_\odot=9.2^{+4.1}_{-7.2}~\kpc$. \citet{Rich2015} confirmed the presence of the HVC in the direction of this star, which places the HVC at a distance of $d_\odot<13.3~\kpc$, and asserted that the HVC lies too far from the LMC to be consistent with an LMC origin. They reason that it is unlikely that material ejected from the LMC would be able to reach a distance of roughly $40~\kpc$ intact during the few hundred million years it would take to travel at speeds of $\sim150~\kms$. This is because, during the cloud's journey it would sustain numerous collisions with not only the CGM of the LMC, but also the halo of the MW. Consequently, \citet{Rich2015} argue that this would strip a large amount of the cloud's material and drag forces would slow the cloud. Needless to say, the time required to make this journey ($250-400~\rm Myr$; \citealt{Barger2016}) would consequently lead to a transverse displacement and the cloud would no longer be toward the LMC.

In light of the results presented in this paper, and by previous studies, we offer a mutual theory on the distribution and association of material observed between the MW and LMC. We postulate that there are two HVCs with different origins near the LMC on the sky: (1) an HVC is associated with the galactic winds of the LMC and (2) an HVC that is associated with the MW. Strong evidence for this latter HVC was presented by \citet{Werner2015} and \citet{Rich2015}, who confirmed that there is high-velocity material at $(l,\,b)=(279\fdg9,\,-37\fdg1)$, $3-5$ degrees away from 30~Doradus, that resides at a distance of $d_\odot<13.3~\kpc$. This sightline is on the periphery of the \ha\ emission that spans across the LMC (see Figure~\ref{fig:EM_total}). Meanwhile, similarities in kinematics, dust depletion, and oxygen abundances strongly indicate that most of the high-velocity material in the direction of the LMC has an LMC origin. Unless the previous evidence gathered from \citet{Staveley2003}, \citet{Lehner2009}, and our study is entirely coincidental, there is likely to be more than one complex toward the LMC. Therefore, we argue that the LMC HVC spans a much larger area of the sky in the direction of the LMC and that there is also a smaller HVC positioned just offset from the LMC on the sky, which is associated with the MW. 

\subsection{Feedback of Low-Mass Galaxies}
\begin{figure}[t!]
\includegraphics[width=0.5\textwidth]{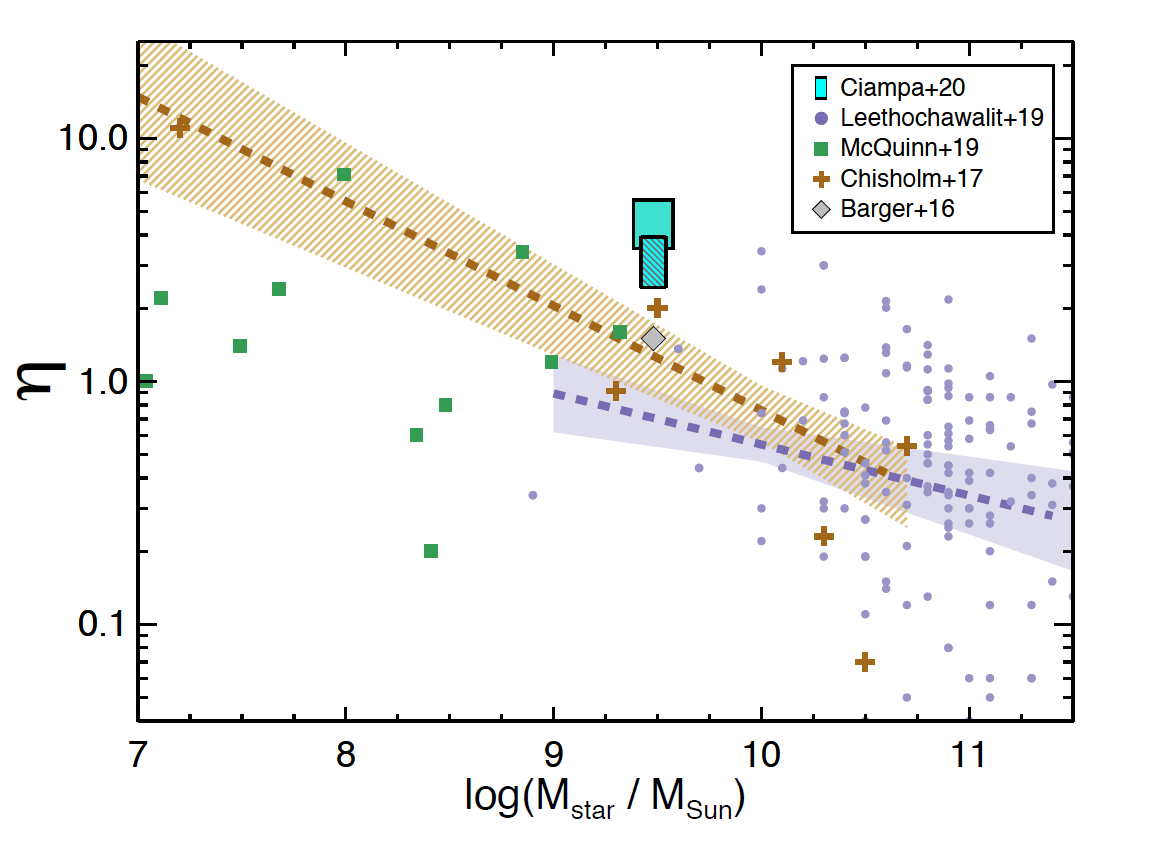}
\caption{Mass-loading factors from various studies. The mass-loading factor we calculated for the total LMC wind (IVC\,+\,HVC) is marked with a cyan bar while solely the IVC component is the darker hashed blue bar. Our results for the LMC wind are found in Table~\ref{tab:mass_results}. The \citet{Barger2016} LMC loading factor (IVC only) is indicated with a gray diamond after we adjust their outflow mass to match the assumptions used in our study. Green squares mark mass loading-factors from \citet{mcquinn2019}, which only include the ionized gas mass in outflows. The brown and purple envelopes are the limits of uncertainty for mass-loading relations from the \citet{chisholm2017} and \citet{lee2019} studies (refer to their equations 16 and 8, respectively).}
\label{fig:mass_loading}
\end{figure}

Galactic winds and the ability of a galaxy to retain ejected gas are directly related to the galaxy's capacity to form future stars. These winds tend to intensify as stellar activity increases. Moreover, because lower mass galaxies have smaller gravitational potential wells, their gas is more easily ejected out of them. For massive galaxies, the halo and extraplanar gas that surrounds their disks suppresses outflowing material. This results in a general trend in which the mass-loading factor is diminished for massive galaxies and enhanced in low-mass galaxies. In the case of the LMC, with a stellar mass of $M_\star=3\times10^9~M_\odot$ \citep{vandermarel2009} and total mass $M_{\rm total}=1.7\times10^{10}~M_\odot$ \citep{vandermarel2014}, as well as an active star-formation history \citep{Harris2009}, it is expected that this galaxy will have a relatively elevated mass-loading factor.

We estimate that the LMC's mass-loading factor ranges from $3.53\leq\eta\leq5.56$ when including both the IVC and HVC gas in its winds and when assuming cylindrical geometry. Our values are relatively large when compared to other studies for galaxies with similar stellar mass (see Figure~\ref{fig:mass_loading}). However, the \citet{Barger2016}, \citet{chisholm2017}, and \citet{lee2019} studies all used absorption-line spectroscopy, which spatially probes less of the wind; this results in a more uncertain mass estimate as the physical extent of winds are poorly constrained. In the case of the \citet{mcquinn2019} \ha\ imaging study, although they were able to measure the extent of the wind, their observations were more than a magnitude less sensitive than ours and were unable to detect the diffuse gas in the wind.

The impact of geometry assumptions is most telling when comparing our results with those of the \citet{Barger2016} absorption-line study of the LMC. Although our study shares many of the same assumptions as their study, we were able to map the extent and morphology of the wind, whereas they were forced to make simplistic assumptions for its solid angle. The mass-flow rates from the \citet{Barger2016} and the star-formation rates from \citet{Harris2009} correspond to a mass-loading factor for LMC that is in the range $0.7\leq\eta\leq0.8$. When we increase their solid angle to match what we observe in \ha\ emission, their mass-loading factor becomes $1.3\leq\eta\leq1.5$. This revised range is in better agreement with our mass-loading factor for conical geometries (Outward and inward cone; see Table~\ref{tab:mass_results}). 

In studies that explore a wide range of galaxy masses, the trend that the mass-loading factor decreases with galaxy mass is clear (see Figure~\ref{fig:mass_loading}). \citet{chisholm2017} find mass-loading factors ranging from $0.2\,-\,19.0$ for their sample of 8~dwarf star-forming galaxies ($6.9\le\log{\left(M_\star/M_\odot\right)} \le10.7$) using UV absorption-line spectroscopy. At the LMC mass, this study predicts a mass-loading factor around $1.1$. Across similar masses, \citet{mcquinn2019} observed their galaxies via \ha\ imaging and found mass-loading factors ranging from $0.2\leq\eta_{\rm ionized}\leq7.1$ for the ionized gas in the wind only. For more massive galaxies ($9.5\le\log{\left(M_\star/M_\odot\right)} \le11.5$), \citet{lee2019} find mass-loading factors that range from $0.3\,-\,1.0$ for their UV absorption-line study. While our largest LMC mass-loading factor estimate is at least twice as large as the \citet{chisholm2017} and \citet{lee2019} trendlines predict, we acknowledge that is in part because of differences in the assumed geometries of the winds. When assuming a conical geometry, a commonly assumed geometry for galactic winds, our mass-loading factor is reduced by nearly 50\% and is more consistent with their results (see Table~\ref{tab:mass_results}).

Still, there is a significant spread in mass-loading factors between studies that requires additional work to reduce. As of right now, no uniform sample exists across low- and high-mass galaxies. Such a consistent sample would go a long way in improving our understand of the relationship between mass-loading factor and galaxy mass. Further, more imaging or emission-line studies are needed to better constrain the geometry of these winds, though they also need to have a high enough sensitivity to detect the diffuse gas. 

\section{Summary} \label{sec:summary}
We completed the first kinematically resolved survey of the LMC's near-side galactic wind in \ha\,$\lambda6563$ using the Wisconsin \ha\ mapper. These mapped observations span 20\,x\,20 degrees across the sky and are comprised of \sightlines sightlines. By combining these observations with existing \hi\ observations, we are able to determine the extent and morphology of the neutral and warm ionized ($T_e\approx10^4~\rm K$) phases of this wind. Here we summarize the main conclusions of this study:

\begin{enumerate}
\item{\bf Morphology and Extent:} We find that diffuse gas in the galactic wind spans across the entire face of the LMC. We additionally find numerous faint $I_{\rm H\alpha}\approx100~\rm mR$ clouds offset from the main wind structure, but we are unable to confidently determine whether or not they are physically associated with the LMC's galactic wind as tidally displaced Magellanic Cloud material and MW HVCs also pollute this region of the sky.

\item{\bf Kinematic Distribution:} We find the bulk of the LMC's galactic wind is moving with velocities of $\rm v_{LMCSR}\lesssim-110~\kms$ relative to the \hi\ disk, which is less than the escape velocity. However, roughly $\log{\left(M_{\rm ion,\,hvc}/M_\odot\right)}\approx7.0$, or roughly 44\%, of this wind is moving away from the LMC at speeds greater than the escape velocity. Specifically, we find the gas that is spatially aligned with 30~Doradus is moving at the greatest speeds relative to the LMC at $\rm v_{LMCSR}\lesssim-175~\kms$.

\item{\bf Two HVC Complexes toward the LMC:} We find \ha\ emission at high velocities relative to the LMC that is strongly spatially correlated with the 30~Doradus. This emission similarly persists at lower velocities. Our results---in addition to the results from the \citet{Staveley2003}, \citet{Lehner2007} \citet{Lehner2009}, and \citet{Barger2016} studies---lead us to conclude that this starburst region is responsible for generating this HVC (see Figures~\ref{fig:spectra} and \ref{fig:CM_ha}). The HVC discussed in \citet{Rich2015}, which lies a few degrees off the $\log{\left(\rm N_{H\textsc{~i}}/\cm^{-2}\right)}=19$ contour of the LMC's \hi\ disk and at a distance of $d_\odot\lesssim13.3~\kpc$, is a different complex. That HVC likely has a MW origin based on their results. 

\item{\bf Outflow Mass, Flow Rate, \& Loading Factor:} We measure an ionized gas mass in the range $7.28\leq\log{\left(M_{\rm ionized}/M_\odot\right)}\leq7.43$ for the outflowing material on the near-side of the LMC that is moving at intermediate-velocities, i.e., speeds that are within $\sim100~\kms$ of the LMC's \hi\ disk. The high-velocity component of this wind has an ionized mass of $6.93\leq\log{\left(M_{\rm ionized}/M_\odot\right)}\leq7.05$. Combined, we estimate that the total ionized gas mass in this near-side wind is in the range $7.44\leq\log{\left(M_{\rm ionized}/M_\odot\right)}\leq7.58$. This corresponds to a total neutral and ionized mass of the entire wind that ranges between $7.87\leq\log{\left(M_{\rm total}/M_\odot\right)}\leq8.0$, assuming that it is symmetrical on the near-side and far-side of the LMC and that it is 75\% ionized (see \citealt{Lehner2007} and \citealt{Barger2016}). We further calculate a total mass-flow rate and mass-loading factor of $1.20\leq\dot{M}_{\rm outflow}\leq1.67~M_\odot~\rm yr^{-1}$ and $3.53\leq\eta\leq5.56$. Table~\ref{tab:mass_results} summarizes these results.

\item{\bf Undetected Diffuse Material:} We compared our results with existing mass-loading factor trends that vary with stellar mass. We find that our average mass-loading factors are on average roughly 2.5~times larger than both optical \ha\ imaging and UV absorption-line studies at the stellar mass of the LMC. This indicates that either the observational sensitivity (optical imaging: \citealt{mcquinn2019}) may be insufficient to detect diffuse gas in these outflows or that the geometric assumptions are too conservative (UV absorption-line spectroscopy: \citealt{Barger2016}, \citealt{chisholm2017}, and \citealt{lee2019}).
\end{enumerate}

\section*{Acknowledgments}
We thank Lister Staveley-Smith and Sungeun Kim for providing us with the ATCA and Parkes telescopes LMC \hi\ survey datacube. This paper includes archived LAB and GASS \hi\ data obtained through the AIfA \hi\ Surveys Data Server (https://www.astro.uni-bonn.de/hisurvey/index.php). WHAM operations for these observations were supported by National Science Foundation (NSF) awards AST-1108911 and AST-1714472/1715623. Madeline Horn received additional support through NSF grant PHY-1358770. 

\bibliography{cite.bib}

\end{document}